\documentclass[a4paper,11pt]{article}
\usepackage{jinstpub}
\usepackage{lineno}
\usepackage{xcolor}
\usepackage{multirow}
\usepackage{comment}
\usepackage{todonotes}
\usepackage{url}
\usepackage{hyperref}

\title{\boldmath Characterization of Next Generation Power Over Fiber Systems under Cryogenic Conditions for Advanced Physics Instrumentation}

\author[a,1]{J. Rodríguez Rondon \note{Corresponding author.}}
\author[a]{D. Leon Silverio,}
\author[a]{H. Wolf,}
\author[a,2]{D. A. Martinez Caicedo,\note{Corresponding author.}}
\author[b]{ and W. Pellico} 

\affiliation[a]{South Dakota School of Mines and Technology,\\ Rapid City, SD 57701, USA}
\affiliation[b]{Fermi National Accelerator Laboratory, \\Batavia, IL, 60510, USA}

\emailAdd{David.MartinezCaicedo@sdsmt.edu, Jairo.RodriguezRondon@sdsmt.edu}

\abstract{The successful application of Power-over-Fiber (PoF) technology to power photon detection systems in  DUNE prototypes operating under high-voltage and cryogenic conditions has laid the groundwork for the next generation of PoF systems operating in harsh environments. This paper presents the characterization of the main components of a new PoF system operating at 1470 and 1550 nm, including the measurement of the electrical-to-optical conversion efficiency of laser transmitters, the optical-to-electrical conversion efficiency of optical power converters (OPCs), and a procedure to estimate optical fiber attenuation at cryogenic conditions. In addition, a detailed characterization of the response of silicon photomultipliers (SiPMs) to light leakage from optical fibers and OPCs is presented, using 808 nm as a reference to evaluate the reduction in the SiPM response at longer wavelengths. The results of this work provide a foundation for the development of next-generation PoF systems capable of powering electronics and electrical systems for future high-energy physics experiments. In addition, the electrical isolation, low-noise operation, and long-term reliability of PoF technology could motivate its application in other research fields requiring remote power delivery under cryogenic conditions, as well as in systems where infrared photons could affect their performance.}


\begin{document}

\begin{tikzpicture}[remember picture,overlay]
    \node[anchor=north east, align=right, font=\small] at (current page.north east) {FERMILAB-PUB-26-0758-LBNF};
\end{tikzpicture}

\maketitle
\flushbottom

\clearpage
\section{Introduction}
\label{sec:intro}

Power-over-Fiber (PoF) technology is used to transmit optical power from a laser via a lightweight, insulating optical fiber to a photovoltaic receiver known as an optical power converter (OPC). This technology has been widely used in systems requiring electrical isolation, particularly in harsh environments where electrical interference needs to be minimized. OPCs convert optical power into electrical power through the photovoltaic effect in semiconductor materials, without moving parts. The passive nature of the OPC offers potential advantages for long-term operation and reliability.

The novel application of PoF technology in the Deep Underground Neutrino Experiment (DUNE) to power the cold electronic readout of DUNE photon detection system (PDS) prototypes marked the first successful application in harsh environments in high energy physics \cite{arroyave2024characterization}. In the DUNE Vertical Drift far detector module (FD-VD), the PDS will be composed of X-ARAPUCA photon detector devices \cite{DUNEFDVD2024,Machado:2018rfb}; some of these modules will be placed on the cathode plane of the detector and readout by a cold electronics module powered via a PoF system operating at 808 nm. A similar need for electrically isolated power delivery may arise in the DUNE third far detector module (FD3), where the proposed Aluminum Profiles with Embedded X-ARAPUCA (APEX) concept \cite{DUNE2024PhaseII,DUNE2025PhaseIIDetectors,Marinho2025APEX} is under consideration to place photon detectors on the liquid argon time projection chamber field cage walls.
 
In both FD-VD and the proposed PDS for FD3, the photon detectors are in close proximity to high voltage environments within the cryostat. Therefore, the use of PoF technology provides an advantage by enabling electrically isolated power delivery to the cold readout electronics of the PDS. Previous results with a 808 nm PoF system used GaAs OPCs with an optical-to-electrical power conversion efficiency in the range of 50-55$\%$. However, certain challenges appeared during the deployment of the 808 nm PoF system in DUNE prototypes including the sensitivity of the DUNE PDS to IR photons escaping the PoF system components, and the power loss associated with optical fibers operating in cryogenic conditions. The need to reduce fiber attenuation losses and minimize IR light leakage will be important for future PoF integration in photon detection systems used in HEP experiments.

Recently, the development of InGaAs OPCs able to operate at 1470 nm, with efficiency up to ~65$\%$~\cite{Fafard2024}, opened up the exploration of these longer wavelengths as viable option for the next generation of a PoF system operating in harsh conditions. Beyond high energy physics, PoF technology has been used when lightweight, electrically isolated connections with low attenuation are desirable. Such applications include optically powered sensor systems or remote antenna units~\cite{MatsuraPhotonics}. Applications of fiber-optic systems in harsh environments, such as aerospace technology~\cite{rovera2023fiber} and quantum computing~\cite{chang2025co}, are under development. While the results presented in this paper serve as foundation to design a next generation PoF system for powering the proposed photon detection system for DUNE FD3, the procedures developed to characterize the PoF components could also be applied in the aforementioned research fields. 

In this paper, the characterization of each component of a next generation 1470 nm and 1550 nm PoF system is presented; the structure will be as follows: Section \ref{sec:lasers} describes the laser modules used to provide power to the PoF system and their electrical-to-optical power efficiency, as well as the performance of a customized laser driver prototype. Section \ref{sec:opticalfibers} discusses the studied optical fibers and the characterization of their attenuation in warm and cryogenic environments. Section \ref{sec:OPC} details the OPCs tests and reports their optical-to-electrical power conversion efficiency. Section \ref{sec:light_leak} describes the characterization of light leakage from the optical fibers and OPCs and its effect on the Broadcom silicon photomultipliers used in this study, emphasizing the comparison with the light leakage measured using an 808~nm PoF system. Section \ref{sec:applications} discusses the potential application of the next generation PoF system in the proposed photon detection system for the third DUNE far detector module as well as other research applications. Finally, Section \ref{sec:summary} provides a summary of the results discussed in this paper.
\section{1470 and 1550~nm laser diodes for PoF}
\label{sec:lasers}

In a PoF system, the laser transmitter acts as the initial connection between electrical and optical power. The electrical power supplied to the laser transmitter is converted to optical power that can be transmitted through an optical fiber. Previous work on a 808~nm PoF system~\cite{arroyave2024characterization} used a Broadcom GaAs laser capable of delivering up to 2~W of optical power. Further developments in PoF technology, such as InGaAs-based OPCs, together with the importance of minimizing photon background sources that could be detected by the DUNE PDS, motivated the investigation of longer operating wavelengths in the measurements presented in this paper. In this section, the characterization of 1470 nm and 1550 nm off-the-shelf laser transmitters will be presented, along with development progress for an associated laser driver prototype. The 1470~nm model provides a nominal output power of 2.5~W, while the 1550~nm model provides 1.5~W. The main specifications of both laser diodes are summarized in Table~\ref{tab:laser_specs}. These lasers provided the optical input power used for the OPC characterization and fiber light-leakage measurements described in Section \ref{sec:OPC} and \ref{sec:light_leak}. The nominal laser output powers reported in the data sheets were not used directly in the analysis. Instead, the optical power of the laser was measured at the laser's pigtail fiber output  for each test and used as the corresponding input power, $P_\mathrm{in}$. Figure~\ref{fig:laser_setup} shows the setup used with the 1470 and 1550~nm laser transmitters. 

\begin{table}[!ht]
  \centering
  \caption{Main specifications of the 1470~nm and 1550~nm fiber-coupled laser transmitters used in this work.}
  \vspace{1em}
  \begin{tabular}{lcc}
    Parameter & 1470~nm laser & 1550~nm laser \\
    \hline
    Nominal output power & 2.5~W & 1.5~W \\
    Central wavelength & 1470~nm & 1550~nm \\
    Wavelength range & $\pm 20$~nm & $\pm 20$~nm \\
    Fiber core diameter & 62.5 $\mu$m & 62.5 $\mu$m \\
    Numerical aperture & 0.22 & 0.22 \\
    Fiber connector & FC/SMA & FC/SMA \\
    Working current & 9.5~A & 6~A \\
    Threshold current & 0.4~A & 0.5~A \\
    Power conversion efficiency & 35--50\% & 20\% \\
    Differential quantum efficiency & 0.7--0.9~W/A & 0.36~W/A \\
    Working voltage & 1.8~V & 1.5~V \\
    Working temperature & 10 to 30~$^\circ$C & 10 to 30~$^\circ$C \\
    Storage temperature & $-20$ to $+60~^\circ$C & $-20$ to $+60~^\circ$C \\
    Life expectancy & $>10000$~h & $>10000$~h \\
  \end{tabular}
  \label{tab:laser_specs}
\end{table}

\subsection{Characterization of 1470 and 1550 nm lasers}
\begin{figure}[!ht]
    \centering
    \includegraphics[width=0.7\linewidth]{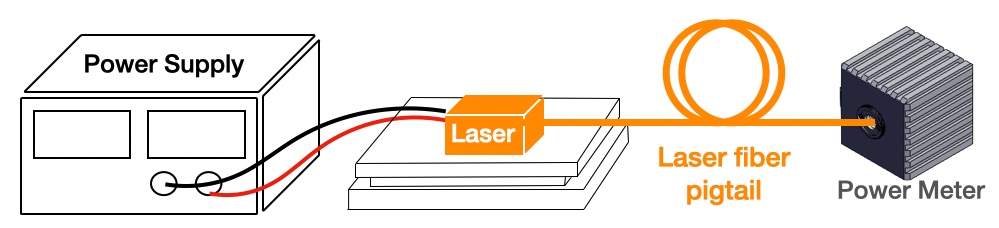}
    \caption{Setup used for the laser output-power measurements. Laser transmitter is mounted on a Peltier-based temperature-control stage and powered by the DC power supply. The laser output is measured with an power sensor.}
    \label{fig:laser_setup}
\end{figure}

The electrical and optical performance of the 1470 and 1550 nm laser transmitters was characterized. The experimental setup used for these measurements is shown in Figure \ref{fig:laser_setup}. The laser transmitter was mounted on a Peltier-based temperature-control plate and connected to a DC power supply. The laser fiber pigtail was connected directly to a Thorlabs S145C integrating-sphere photodiode power sensor \cite{Thorlabs_S145C} to measure the optical output power.

\begin{figure}[!h]
    \centering
    \includegraphics[width=0.48\linewidth]{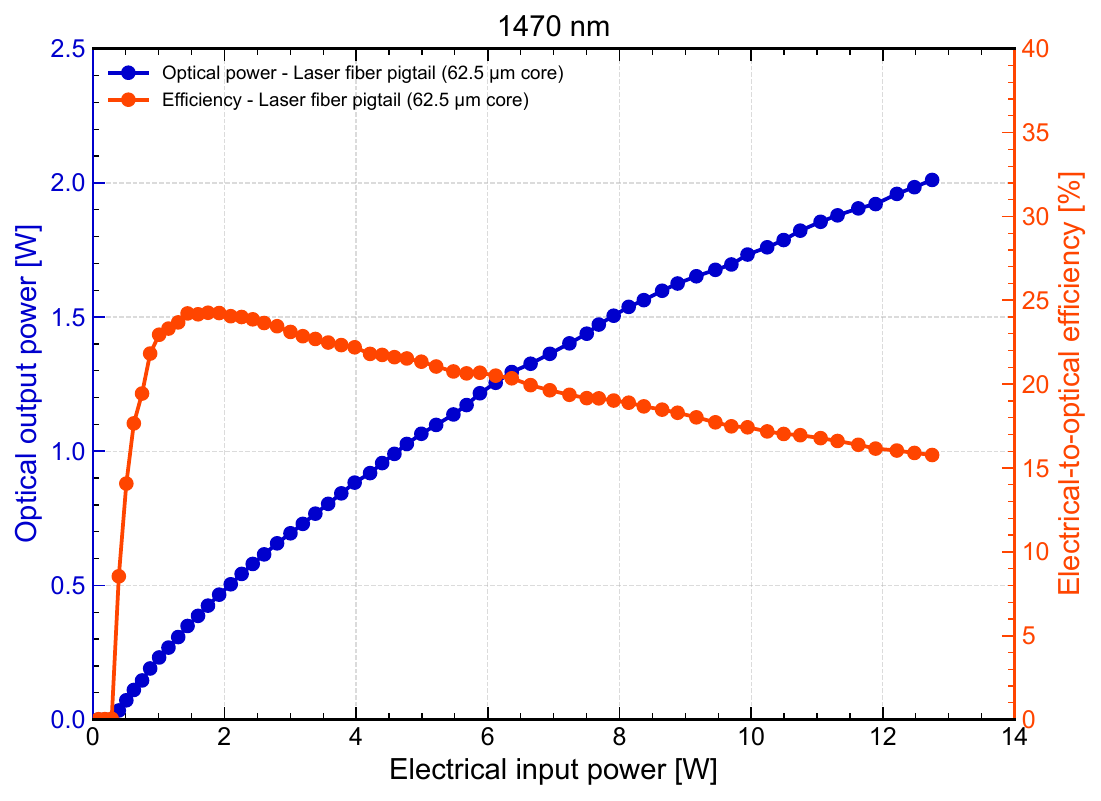}
    \includegraphics[width=0.48\linewidth]{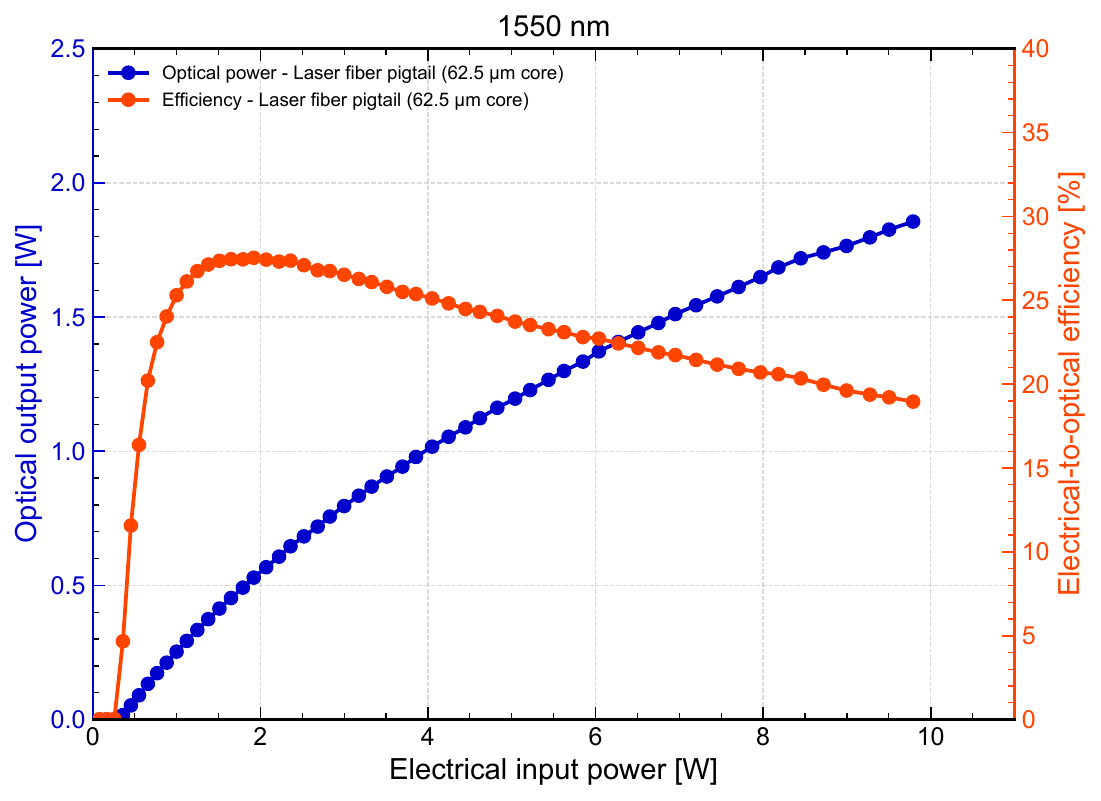}
    
    \caption{Electrical-to-optical characterization of the 1470 and 1550 nm laser transmitters using a 62.5 $\mu$m-core pigtail. The optical output power (blue) and electrical-to-optical conversion efficiency (orange) are shown as functions of the electrical input power for the 1470 nm transmitter (left) and the 1550 nm transmitter (right).}
    \label{fig:laser_characterization}
\end{figure}

The laser transmitter characterization was performed by progressively increasing the current supplied to the laser using the DC power supply. At each operating point, the voltage across the laser and the corresponding optical output power at the pigtail fiber output were recorded. For each operating point, the electrical input power was calculated as:

\begin{equation}
    P_{\mathrm{elec}} = VI,
\end{equation}

where \textit{V} is the measured voltage across the laser and \textit{I} is the supplied current. The electrical-to-optical conversion efficiency of each laser was calculated as:

\begin{equation}
    \eta_{\mathrm{laser}}
    =
    \frac{P_{\mathrm{opt}}}{P_{\mathrm{elec}}}.
\end{equation}

where $P_\mathrm{opt}$ is the measured optical output power and $P_\mathrm{elec}$ is the electrical power supplied to the laser. Figure~\ref{fig:laser_characterization} shows the optical output power and the efficiency ($\eta_{\mathrm{laser}}$) as functions of the electrical input power for the 1470 and 1550~nm lasers. The output power increases with electrical input power for all of the laser transmitters. In contrast, the electrical-to-optical power conversion efficiency rises rapidly, reaches a maximum, and gradually decreases as the electrical input power increases.

For both wavelengths, the nominal electrical-to-optical conversion efficiency occurs at an intermediate electrical input power rather than at the operating point corresponding to the maximum optical output power. Differences in the maximum efficiency are observed between the 1470 and 1550 nm transmitters. For the 62.5 $\mu$m-core pigtail, the 1550 nm transmitter reaches a maximum measured conversion efficiency of approximately 28\% at an electrical input power of 1.92 W, while the 1470 nm transmitter reaches approximately 24\% at 1.75 W. The differences between the measured efficiencies and the values reported in the manufacturer's datasheets are associated with the operating conditions of our experimental test stand, including temperature variations that can affect the laser conversion efficiency.

These measurements show that the maximum optical output power and the maximum electrical-to-optical conversion efficiency are reached at different operating points. Therefore, the laser operating current can be selected based on the optical power required by the PoF system while considering the corresponding electrical-to-optical conversion efficiency.

\subsection{Prototyping laser drivers for 1470 and 1550 nm laser transmitters}

Preliminary results of a laser driver prototype performance that could serve for the 1470 and 1550 nm laser transmitters are presented in this section. The 1470 and 1550~nm laser transmitters considered in this work can require currents up to approximately 9 and 6~A, respectively, at their nominal operating conditions, motivating the development of a dedicated high-current driver solution. 

\begin{figure}[!ht]
    \centering
    \includegraphics[width=0.46\linewidth]{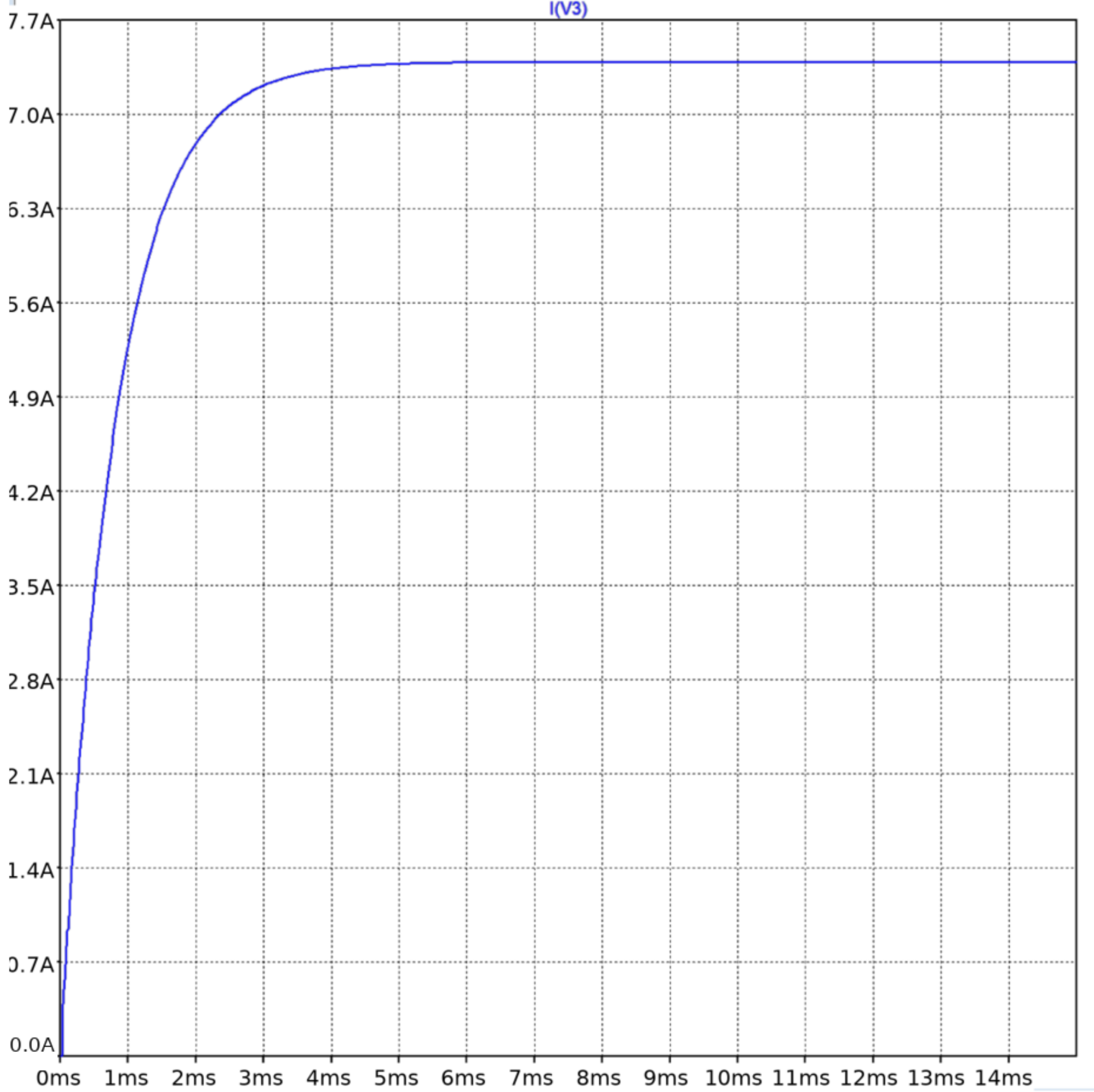}
    \includegraphics[width=0.48\linewidth]{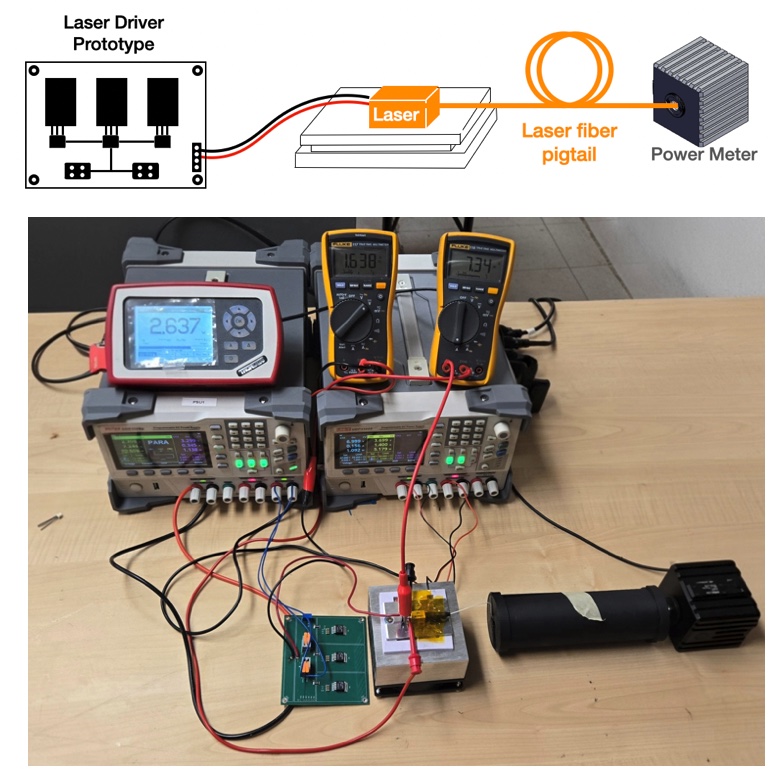}
    \caption{Laser-driver prototype and its validation through simulation and experimental testing. Left: shows the simulated current response of the driver, reaching approximately 7.3~A. Right: Shows a schematic of the experimental setup and  the assembled test setup, where the laser transmitter was mounted on a Peltier-based temperature-control plate and connected to the laser driver prototype, with the optical output measured using an optical power meter. In the configuration shown, the driver operated at approximately 7.3~A ($\sim$15 W electrical input power) while the measured optical output was about 2.6~W for a 1470 nm laser.}
    \label{fig:laser_driver_p}
\end{figure}

The current prototype was designed to provide up to approximately 7~A, successfully reaching this current during prototype testing. Figure~\ref{fig:laser_driver_p} (left) shows the simulated current response of the designed laser driver, which stabilizes at approximately 7.3 A. The experimental set up to characterize and validate the laser driver prototype is shown in Figure \ref{fig:laser_driver_p} (right). The laser driver prototype was connected to the laser transmitter mounted on the Peltier-based temperature-control plate, while the optical output was measured with a power meter.

\begin{figure}[!h]
    \centering
    \includegraphics[width=0.89\linewidth]{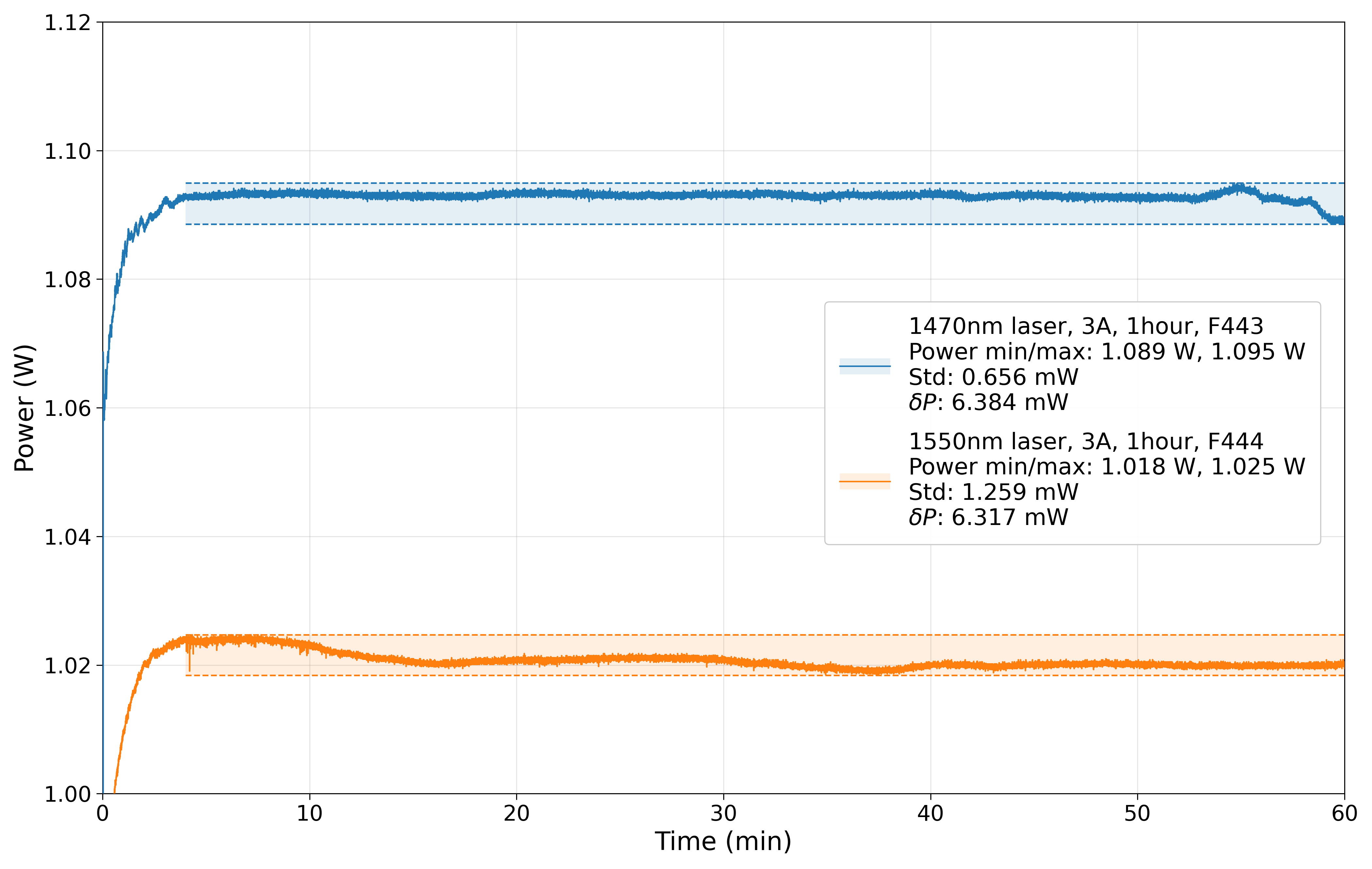}
    \caption{Optical power output of the 1470~nm (blue) and 1550~nm (orange) laser transmitters during a one-hour stability test. Both lasers were operated at 3~A supplied by the driver prototype. After a short warm-up period, the stable optical output of both lasers indicates that the prototype provided steady current regulation throughout the test, with power fluctuations around 1\%.}
    \label{fig:laser_driver_stability}
\end{figure}

The prototype has already been tested with both the 1470 and 1550~nm laser diodes. As an initial power stability test, both lasers were operated continuously at 3~A for one hour while monitoring their optical output power using a power sensor. As shown in Figure~\ref{fig:laser_driver_stability}, after a short warm-up period, both lasers maintained a stable output with power fluctuations around 1\%. These results demonstrate a preliminary validation of the laser driver prototype for extended operation, while further optimization and longer-term tests are planned, aiming to be integrated with future large-area photon detectors proposed for DUNE Phase II \cite{Marinho2025APEX,DUNE2024PhaseII}. For the measurements presented in section \ref{sec:opticalfibers}, \ref{sec:OPC} and \ref{sec:light_leak}, the laser transmitters were operated without the laser driver prototype.

\section{Optical fibers}
\label{sec:opticalfibers}

Optical fibers have been widely used for several decades, particularly in telecommunications, where fibers allow the transmission of optical signals over long distances with relatively low losses. Optical fibers applications have expanded to other fields, including sensing, medical instrumentation, and scientific experiments operating in challenging environments \cite{lee2001properties, lecocq2021control, li2020study}. In addition to signal transmission, the optical fibers can be also used to deliver optical power, as implemented in the PoF systems. The optical power transmitted through the fiber depends on the fiber characteristics, operating wavelength, and environmental conditions. For applications at cryogenic temperatures, it is important to characterize the optical power losses to evaluate the transmission performance of the fibers under these conditions.

Previous studies performed for the DUNE FD-VD PDS focused on the characterization and validation of multi-mode optical fibers operating with 808 nm PoF systems \cite{arroyave2024characterization}. In this section, the characterization is extended to 1470 and 1550 nm, considering their potential application in the next generation of PoF systems that could be used in photon detectors, while minimizing the impact of light leakage from the fibers on the photon detector response.

\begin{figure}[h!]
    \centering
    \includegraphics[width=0.7\linewidth]{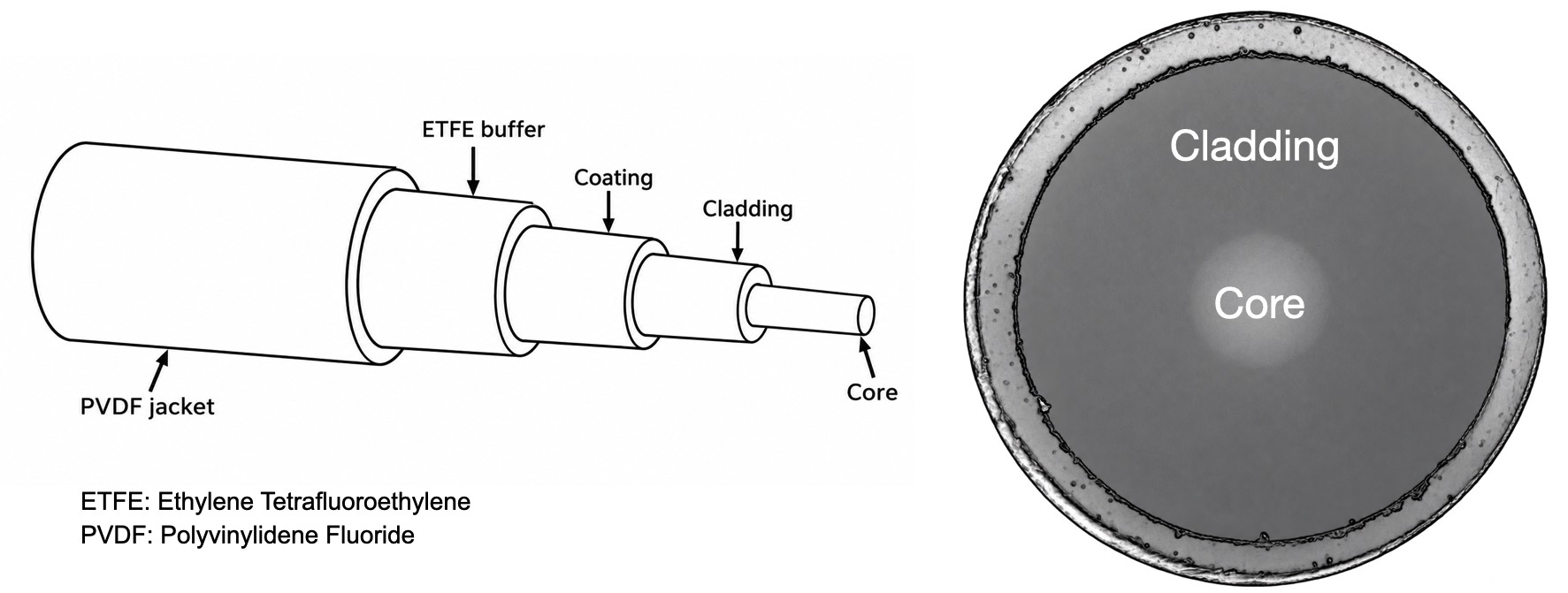}
    \caption{Schematic representation of the optical fiber showing its different layers (left). The cross-sectional image (right), obtained using a Thorlabs optical fiber inspection scope, shows the core and cladding.}
    \label{fig:fiber_schematic}
\end{figure}

The optical fibers used in this study are multi-mode fibers from MH-GoPower. Table \ref{tab:fiber_specs} summarizes the specifications of the fibers, while Figure \ref{fig:fiber_schematic} shows a schematic representation of the fiber structure and an image of its cross-sectional view.

\begin{table}[h!]
  \centering
  \caption{Main specifications of the MH GoPower optical fibers.}
  \vspace{1em}
  \begin{tabular}{lc}
    Parameter & Value \\
    \hline
    Numerical aperture (NA) & 0.27 \\
    Index profile & Graded Index (GI) \\
    Core diameter & 62.5 $\pm$ 3~$\mu$m \\
    Cladding diameter & 200 $\pm$ 4~$\mu$m \\
    Coating diameter & 230 $\pm$ 10~$\mu$m \\
    ETFE buffer diameter & 500 $\pm$ 50~$\mu$m \\
    PVDF jacket diameter & 1.5 $\pm$ 0.1~mm \\
    Connector type & FC/PC Ceramic \\
  \end{tabular}
  \label{tab:fiber_specs}
\end{table}

The optical power transmitted through an optical fiber is affected by losses associated with absorption and scattering. These losses depend on the operating wavelength and the characteristics of the fiber. Operation at cryogenic temperatures can introduce additional optical power losses. To characterize the optical power losses of the fiber under test at 1470 and 1550 nm, measurements were performed at room temperature and liquid nitrogen (LN$_{2}$). The room-temperature measurements were used to determine the optical power loss as a function of fiber length, while the measurements in LN$_{2}$ were used to quantify the additional losses introduced by cryogenic conditions.

\subsection{Room-temperature fiber attenuation}

The optical power loss at room temperature was measured using optical fibers with lengths of 5, 10, and 30 m and two laser transmitters operating at 1470 and 1550 nm. First, the optical power from each laser transmitter was measured using an optical power meter to establish a reference measurement. Then, the fiber under test was connected between the laser transmitter and the power meter to measure the optical power. This procedure was repeated for the three fiber lengths (5, 10 and 30 m) at both wavelengths.

\begin{figure}[h!]
    \centering
    \includegraphics[width=0.9\linewidth]{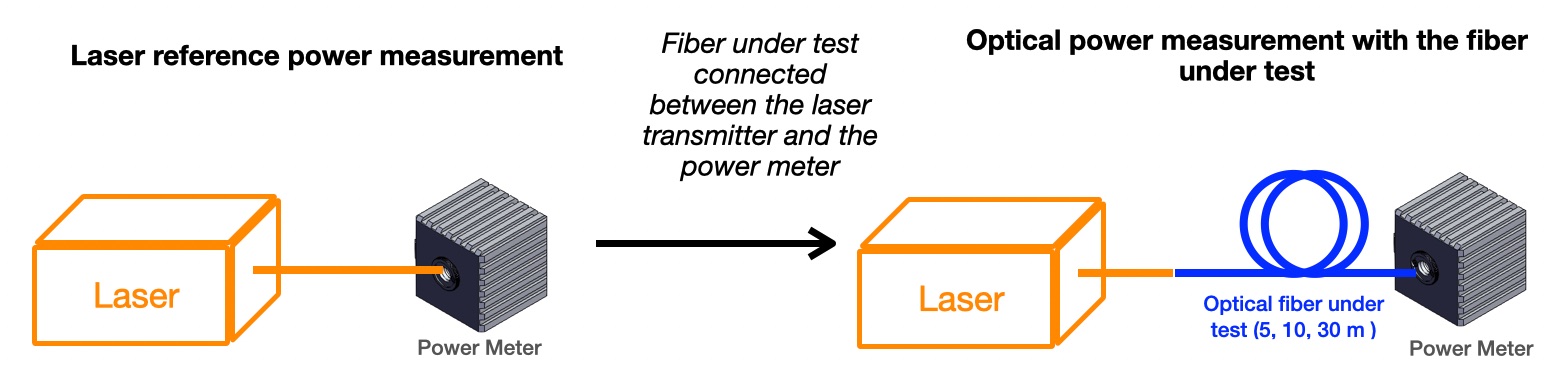}    \caption{Schematic of the experimental setup used to measure optical power losses at room temperature. The reference optical power was first measured directly from the laser transmitter. The fiber under test (5, 10, and 30 m) was then connected between the laser transmitter and the power meter to measure the optical power at the fiber output.}
    \label{fig:roomloss}
\end{figure}

These measurements were used to characterize the attenuation of the optical path, which includes the optical fiber and its FC-to-FC connections. Therefore, the measured optical power losses include contributions from both the fiber transmission and the optical connections. Since the measurements were performed using three independently terminated optical fibers, variations in the FC/PC connector insertion losses and optical coupling between individual fiber assemblies may also contribute to the attenuation coefficient obtained from the linear fit. Figure \ref{fig:roomloss} shows the experimental setup used for these measurements.

For each fiber length, the optical transmission was calculated relative to the corresponding laser reference measurement as:

\begin{equation}
    T_{\mathrm{warm}}(L)
    =
    \frac{P_{\mathrm{fiber}}(L)}
         {P_{\mathrm{ref}}(L)},
    \label{eq:T_warm}
\end{equation}

where $P_{\mathrm{ref}}$ is the optical power measured directly from the laser transmitter and $P_{\mathrm{fiber}}$ is the optical power measured at the output of the fiber of length (L). The room-temperature attenuation is expressed in decibels as:

\begin{equation}
    A_{\mathrm{warm}}(L)
    =
    -10\log_{10}
    \left[
    T_{\mathrm{warm}}(L)
    \right]
    =
    -10\log_{10}
    \left[
    \frac{P_{\mathrm{fiber}}(L)}
         {P_{\mathrm{ref}}(L)}
    \right].
    \label{eq:A_warm}
\end{equation}

The measured transmission includes both the attenuation associated with the fiber length and a length-independent contribution from the optical path. This length-independent contribution is referred to as the fixed insertion loss and accounts for losses associated with the FC-to-FC connections and coupling. The total attenuation of the optical path can therefore be expressed as:

\begin{equation}
    T_{\mathrm{warm}}(L)
    =
    T_{0}T(L)
    =
    T_{0}e^{-\alpha_{\mathrm{fiber}}L},
    \label{eq:T_warm_exp}
\end{equation}

where $T_0$ is the transmission factor associated with the fixed insertion loss and $\alpha_{\mathrm{fiber}}$ is the fiber attenuation coefficient in inverse-length units. Substituting Eq. ~\ref{eq:T_warm_exp} into ~\ref{eq:A_warm}, the attenuation can be written as:

\begin{equation}
    A_{\mathrm{warm}}(L)
    =
    -10\log_{10}(T_{0})
    +
    \frac{10}{\ln 10}\alpha_{\mathrm{fiber}}L
    =
    A_{0}
    +
    4.343\,\alpha_{\mathrm{fiber}}L
    =
    A_{0}
    +
    \alpha_{\mathrm{fiber,dB}}L,
    \label{eq:A_warm_fit}
\end{equation}

where $A_0=-10\log_{10}(T_0)$ is the fixed insertion loss in dB and $\alpha_{\mathrm{fiber,dB}}=4.343\alpha_{\mathrm{fiber}}$ is the room-temperature fiber attenuation coefficient in dB per unit length. The statistical uncertainty in the room-temperature conditions was obtained from the measured power uncertainties as:

\begin{equation}
    \sigma_{A_{\mathrm{warm}}}
    =
    4.343
    \sqrt{
    \left(
    \frac{\sigma_{\mathrm{fiber}}}
         {P_{\mathrm{fiber}}}
    \right)^2
    +
    \left(
    \frac{\sigma_{\mathrm{ref}}}
         {P_{\mathrm{ref}}}
    \right)^2
    }.
    \label{eq:A_warm_uncertainty}
\end{equation}

where $\sigma_{\mathrm{fiber}}$ and $\sigma_{\mathrm{ref}}$ are the standard deviations of the measured optical powers $P_{\mathrm{fiber}}$ and $P_{\mathrm{ref}}$, respectively. The attenuation values obtained for the 5, 10, and 30~m fiber lengths were fitted using ROOT \cite{Brun1997ROOT}. The parameters \(A_{0}\) and \(\alpha_{\mathrm{fiber,dB}}\), together with their fit uncertainties, were obtained directly from the minimization of differences between the measured attenuation and the model 
$A_{\mathrm{warm},i} - \left( A_{0} + \alpha_{\mathrm{fiber,dB}}L_i \right).$

\begin{figure}[h!]
    \centering
    \includegraphics[width=\linewidth]{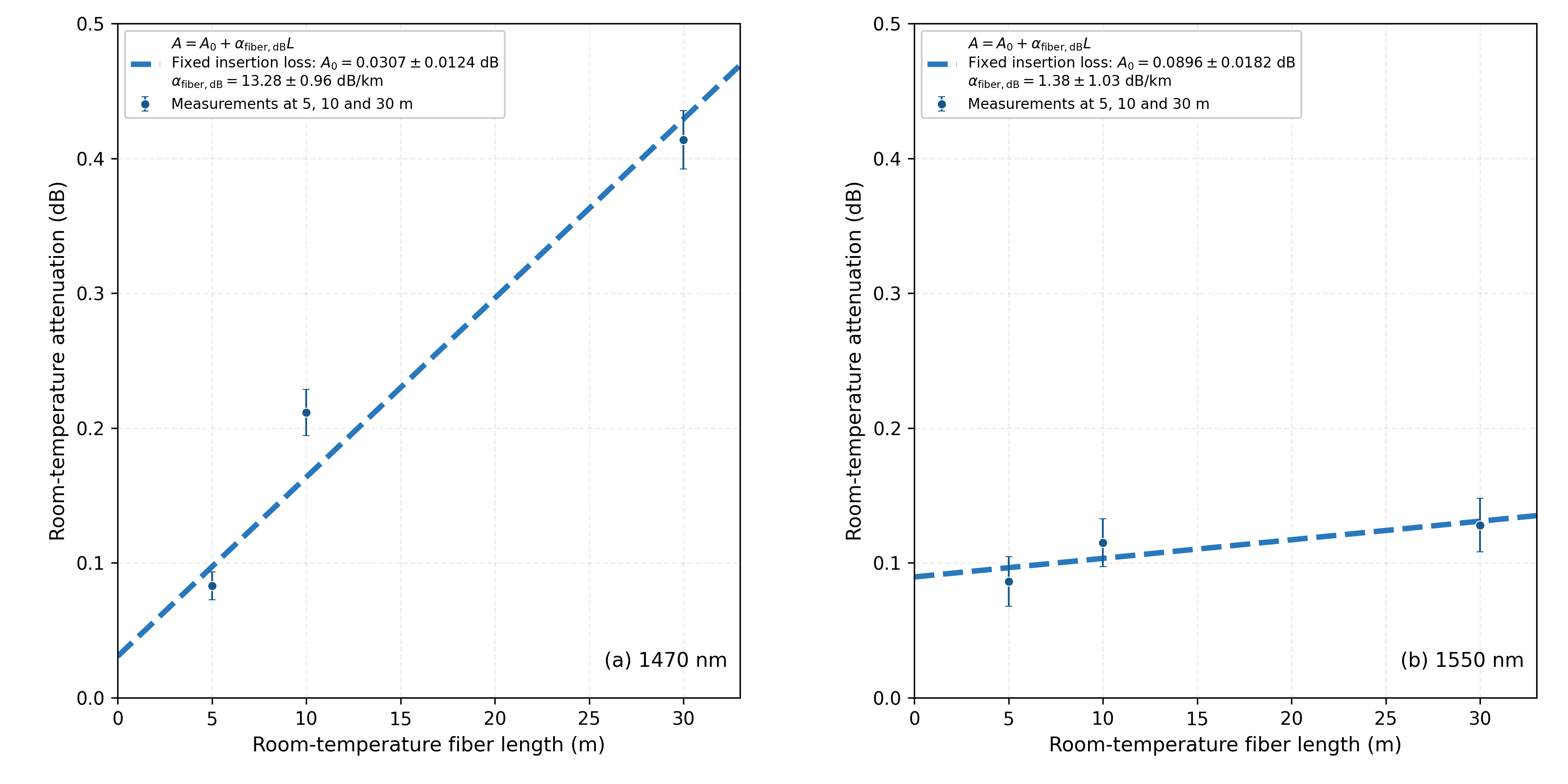}
    \caption{Room-temperature fiber attenuation measured at 1470 and 1550~nm using fiber lengths of 5, 10, and 30~m, shown in panels (a) and (b), respectively. The dashed lines represent the linear fits $A_{\mathrm{warm}}(L)=A_0+\alpha_{\mathrm{fiber,dB}}L$, where $A_0$ is the fixed insertion loss and $\alpha_{\mathrm{fiber,dB}}$ is the room-temperature fiber attenuation coefficient.}
    \label{fig:warm_attenuation_fit}
\end{figure}

Figure~\ref{fig:warm_attenuation_fit} shows the room-temperature attenuation measured as a function of fiber length for the operating wavelengths of 1470 and 1550 nm, together with their corresponding linear fits.

For each wavelength, the fixed insertion loss, $A_0$, and the fiber attenuation coefficient, $\alpha_{\mathrm{fiber,dB}}$, were obtained from the intercept and slope of the fit, respectively. At 1470 nm, the fit gives $A_0=(0.03\pm0.01)$
dB and $\alpha_{\mathrm{fiber,dB}}=(13.28\pm0.96)$ dB/km. At 1550 nm, the corresponding values are $A_0=(0.09\pm0.02)$ dB and 
$\alpha_{\mathrm{fiber,dB}}=(1.38\pm1.03)$ dB/km. The measurements show a lower room-temperature fiber attenuation coefficient at 1550 nm compared with 1470 nm. These parameters were used to estimate the optical losses for longer fiber lengths. 
However, the attenuation measured at 1470 nm is considerably higher than the losses commonly reported for telecommunication optical fibers operating at near-infrared wavelengths (typically on the order of 1 dB/km or lower). Standard multimode fibers are not always designed or characterized for operation at 1470 nm, and their performance can be affected by water-peak absorption, modal dispersion, and mode-dependent losses associated with bending and changes in light propagation. Therefore, for longer distances a careful fiber selection should consider these aspects in more detail.

\subsection{Induced cryogenic attenuation}

The additional optical power loss produced by submersion in LN$_{2}$ was evaluated using measurements at three optical-power settings of 0.5, 0.75, and 1.0 W. First, the optical power from the laser transmitter was monitored for 15 min using an optical power meter to establish the reference measurement. The fiber was then connected to the laser transmitter, and the optical power at the fiber output was measured at room temperature.

\begin{figure}[h!]
    \centering
    \includegraphics[width=\linewidth]{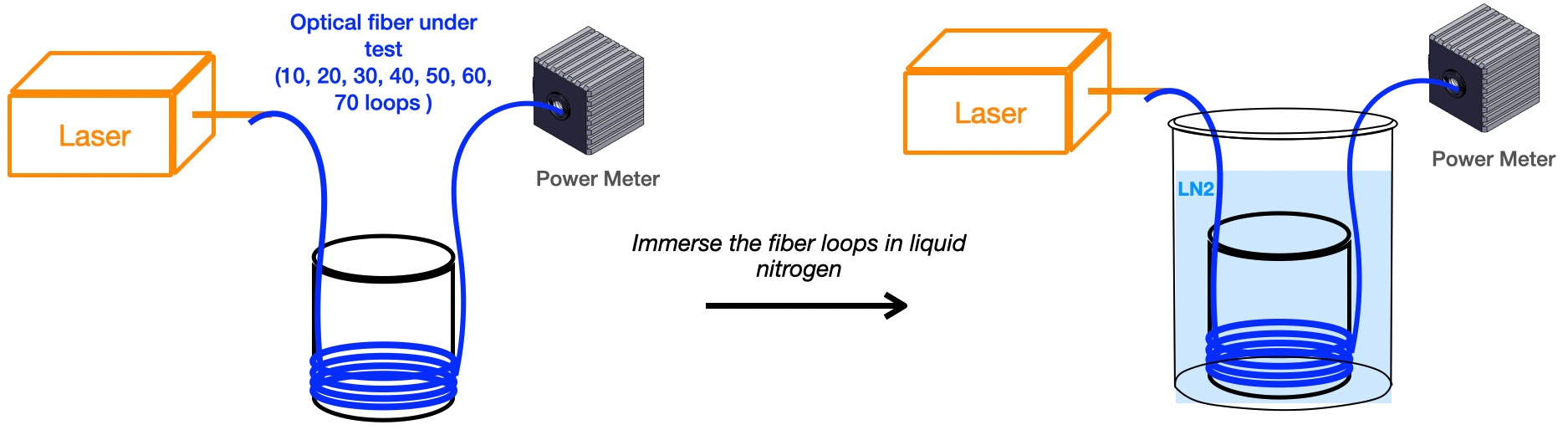}
    \caption{Schematic representation of the experimental setup used to measure the induced cryogenic attenuation of the optical fibers. The fiber under test was wrapped around a cylindrical support with a radius of 6 cm, with the number of loops varying from 10 to 70. The optical power was first measured at room temperature, and the fiber loops were then submerged in LN$_{2}$ to measure the additional optical power losses under LN$_{2}$ conditions.}
    \label{fig:LN2Bending}
\end{figure}

To evaluate the losses at cryogenic temperatures, the optical fiber was wrapped around a cylindrical support with a radius of 6 cm. The number of turns was increased from 10 to 70 in steps of 10 turns, corresponding to submerged fiber lengths of 3.77, 7.54, 11.31, 15.08, 18.85, 22.62, and 26.39 m. The wrapped section of the fiber was submerged in LN$_{2}$, while the optical power at the fiber output was measured using the power meter. This procedure was repeated for the three optical-power settings to evaluate the additional optical power loss as a function of the submerged fiber length. Figure \ref{fig:LN2Bending} shows a schematic representation of the experimental setup used for these measurements.

The optical loss due to the fiber submersion in liquid nitrogen was evaluated relative to the room-temperature power measured through the same fiber assembly. For each optical-power setting, the room-temperature measurement defines the optical power before introducing the cryogenic contribution, $P_{\mathrm{room}}$. When a fiber length (L) is submerged in LN$_{2}$, the corresponding measured power is $P_{\mathrm{LN2}}$. The relative cryogenic transmission is defined as:

\begin{equation}
    T_{\mathrm{cryo}}(L)
    =
    \frac{P_{\mathrm{LN2}}(L)}
         {P_{\mathrm{room}}}.
    \label{eq:T_cryo}
\end{equation}

Using the room-temperature power as the reference removes the room-temperature fiber attenuation and fixed insertion loss that are common to both measurements. Therefore, \(T_{\mathrm{cryo}}(L)\) represents only the additional transmission change associated with submersion in LN$_2$. The corresponding induced cryogenic attenuation is:

\begin{equation}
    A_{\mathrm{cryo}}(L)
    =
    -10\log_{10}
    \left[
    T_{\mathrm{cryo}}(L)
    \right]
    =
    -10\log_{10}
    \left[
    \frac{P_{\mathrm{LN2}}(L)}
         {P_{\mathrm{room}}}
    \right].
    \label{eq:A_cryo}
\end{equation}

The induced loss per unit length of submerged fiber is estimated as:
\begin{equation}
    A_{\mathrm{cryo}}(L)
    =
    \alpha_{\mathrm{cryo,dB}}L,
    \label{eq:A_cryo_fit}
\end{equation}

where \(\alpha_{\mathrm{cryo,dB}}\) is the induced cryogenic attenuation coefficient in dB per unit length. The statistical uncertainty in the induced cryogenic attenuation was obtained from the measured power uncertainties as:

\begin{equation}
    \sigma_{A_{\mathrm{cryo}}}
    =
    4.343
    \sqrt{
    \left(
    \frac{\sigma_{\mathrm{LN2}}}
         {P_{\mathrm{LN2}}}
    \right)^2
    +
    \left(
    \frac{\sigma_{\mathrm{room}}}
         {P_{\mathrm{room}}}
    \right)^2
    }.
    \label{eq:A_cryo_uncertainty}
\end{equation}

where $\sigma_{\mathrm{LN2}}$ and $\sigma_{\mathrm{room}}$ are the standard deviations of the measured optical powers $P_{\mathrm{LN2}}$ and $P_{\mathrm{room}}$, respectively. The attenuation values from the three optical-power settings were fitted simultaneously in ROOT as a function of submerged fiber length. The coefficient $\alpha_{\mathrm{cryo,dB}}$ and its statistical uncertainty were obtained directly from the minimization of the differences between the measured attenuation and the model $\alpha_{\mathrm{cryo,dB}}L$. The fit was constrained through the origin because the induced cryogenic attenuation is zero when no fiber length is submerged.

\begin{figure}[h!]
    \centering
    \includegraphics[width=1\linewidth]{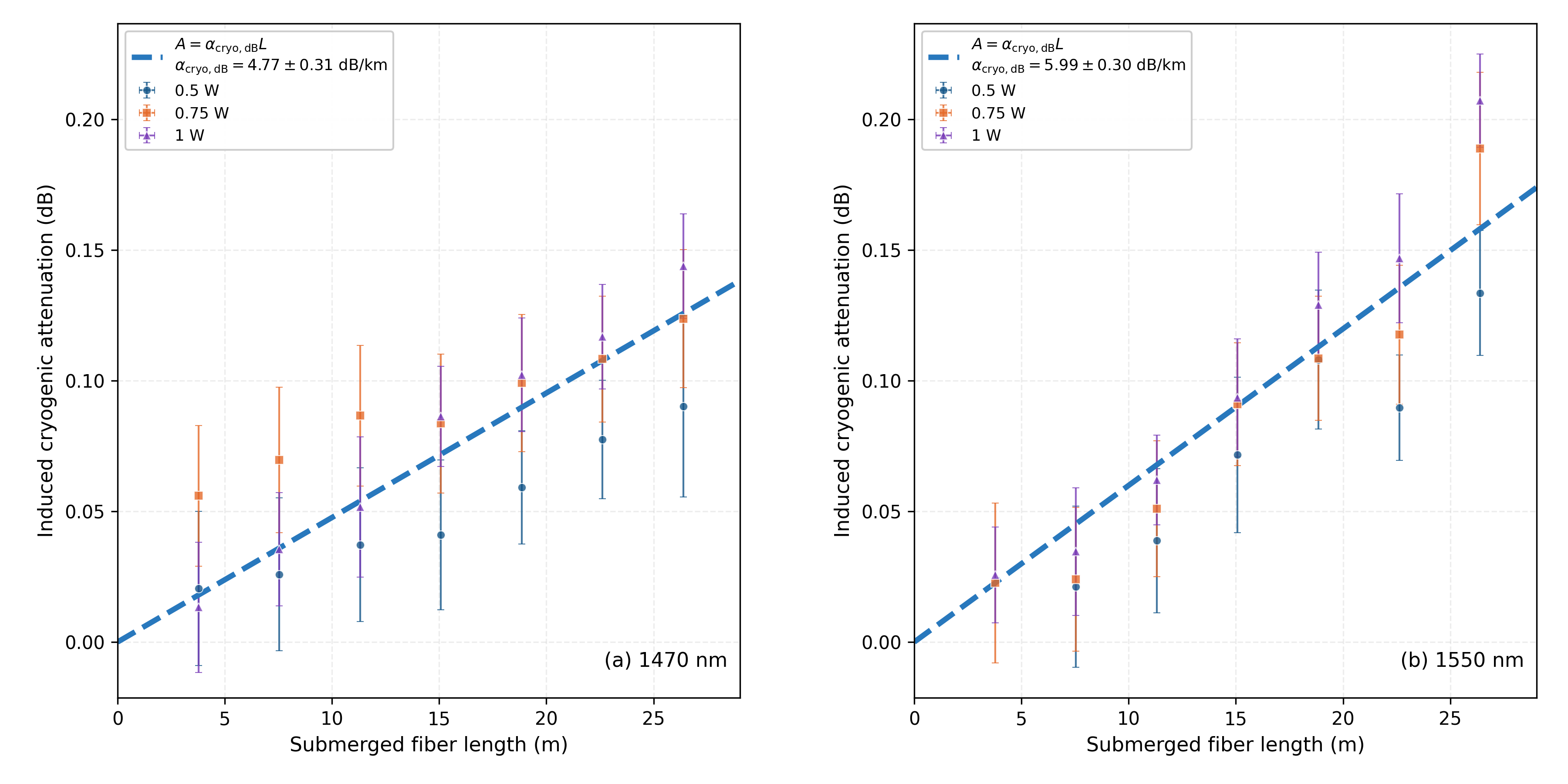}
    \caption{Induced cryogenic attenuation as a function of submerged fiber length at (a) 1470~nm and (b) 1550~nm for three optical-power settings. The dashed lines represent the linear fits $A_{\mathrm{cryo}}(L)=\alpha_{\mathrm{cryo,dB}}L$, constrained through the origin, where $\alpha_{\mathrm{cryo,dB}}$ is the induced cryogenic attenuation coefficient.}
    \label{fig:cryo_attenuation_fit}
\end{figure}

Figure \ref{fig:cryo_attenuation_fit} shows the induced cryogenic attenuation
as a function of submerged fiber length at 1470 and 1550 nm for the three optical-power settings, together with the corresponding linear fits. For each wavelength, the induced cryogenic attenuation coefficient, $\alpha_{\mathrm{cryo,dB}}$, was obtained from the slope of the fit.

At 1470 nm, the fit gives $\alpha_{\mathrm{cryo,dB}}=(4.77\pm0.31)$ dB/km, while at 1550 nm, the corresponding value is $\alpha_{\mathrm{cryo,dB}}=(5.99\pm0.30)$ dB/km. These coefficients were used to estimate the total optical power losses for longer fiber lengths at both wavelengths.

\subsection{Estimation of optical power losses for longer fiber lengths}

The room-temperature and induced cryogenic attenuation measurements were combined to estimate the total optical power losses for longer fiber lengths operating at cryogenic temperatures. The analysis was performed separately for the 1470 and 1550 nm wavelengths using the attenuation coefficients obtained from the measurements presented in the previous sections.

The total attenuation as a function of fiber length, $L$, is expressed as:

\begin{equation}
    A_{\mathrm{total}}(L)
    =
    A_{0}
    +
    \left(
    \alpha_{\mathrm{fiber,dB}}
    +
    \alpha_{\mathrm{cryo,dB}}
    \right)L.
    \label{eq:A_total_projection}
\end{equation}
where $A_0$ is the fixed insertion loss obtained from the room-temperature fit, while $\alpha_{\mathrm{fiber,dB}}$ and $\alpha_{\mathrm{cryo,dB}}$ are the room-temperature and induced cryogenic attenuation coefficients, respectively. Defining the combined attenuation coefficient as $\alpha_{\mathrm{combined,dB}}=\alpha_{\mathrm{fiber,dB}}+\alpha_{\mathrm{cryo,dB}}$, the total attenuation becomes:

\begin{equation}
    A_{\mathrm{total}}(L)
    =
    A_{0}
    +
    \alpha_{\mathrm{combined,dB}}L,
    \label{eq:A_total_compact}
\end{equation}

Table \ref{tab:attenuation_coefficients} summarizes the
room-temperature and induced cryogenic attenuation coefficients obtained for both wavelengths, together with the corresponding fixed insertion losses and combined attenuation coefficients.

\begin{table}[h!]
    \centering
    \caption{Summary of the attenuation parameters obtained
    at 1470 and 1550 nm.}
    \vspace{1em}\label{tab:attenuation_coefficients}
    \begin{tabular}{lcc}
        Parameter & 1470 nm & 1550 nm \\
        \hline
        $A_0$ [dB]
        & $0.03 \pm 0.01$
        & $0.09 \pm 0.02$ \\
        $\alpha_{\mathrm{fiber,dB}}$ [dB/km]
        & $13.28 \pm 0.96$
        & $1.38 \pm 1.03$ \\
        $\alpha_{\mathrm{cryo,dB}}$ [dB/km]
        & $4.77 \pm 0.31$
        & $5.99 \pm 0.30$ \\
        $\alpha_{\mathrm{combined,dB}}$ [dB/km]
        & $18.05 \pm 1.01$
        & $7.37 \pm 1.07$ \\
    \end{tabular}
\end{table}

Using the total attenuation, the fraction of optical power loss as a function of fiber length is calculated as:

\begin{equation}
    f_{\mathrm{loss}}(L)
    =
    1-
    \frac{P_{\mathrm{out}}(L)}{P_{\mathrm{ref}}}
    =
    1-
    10^{-A_{\mathrm{total}}(L)/10}.
    \label{eq:fractional_loss_projection}
\end{equation}

\begin{figure}[h!]
    \centering
    \includegraphics[width=1\linewidth]{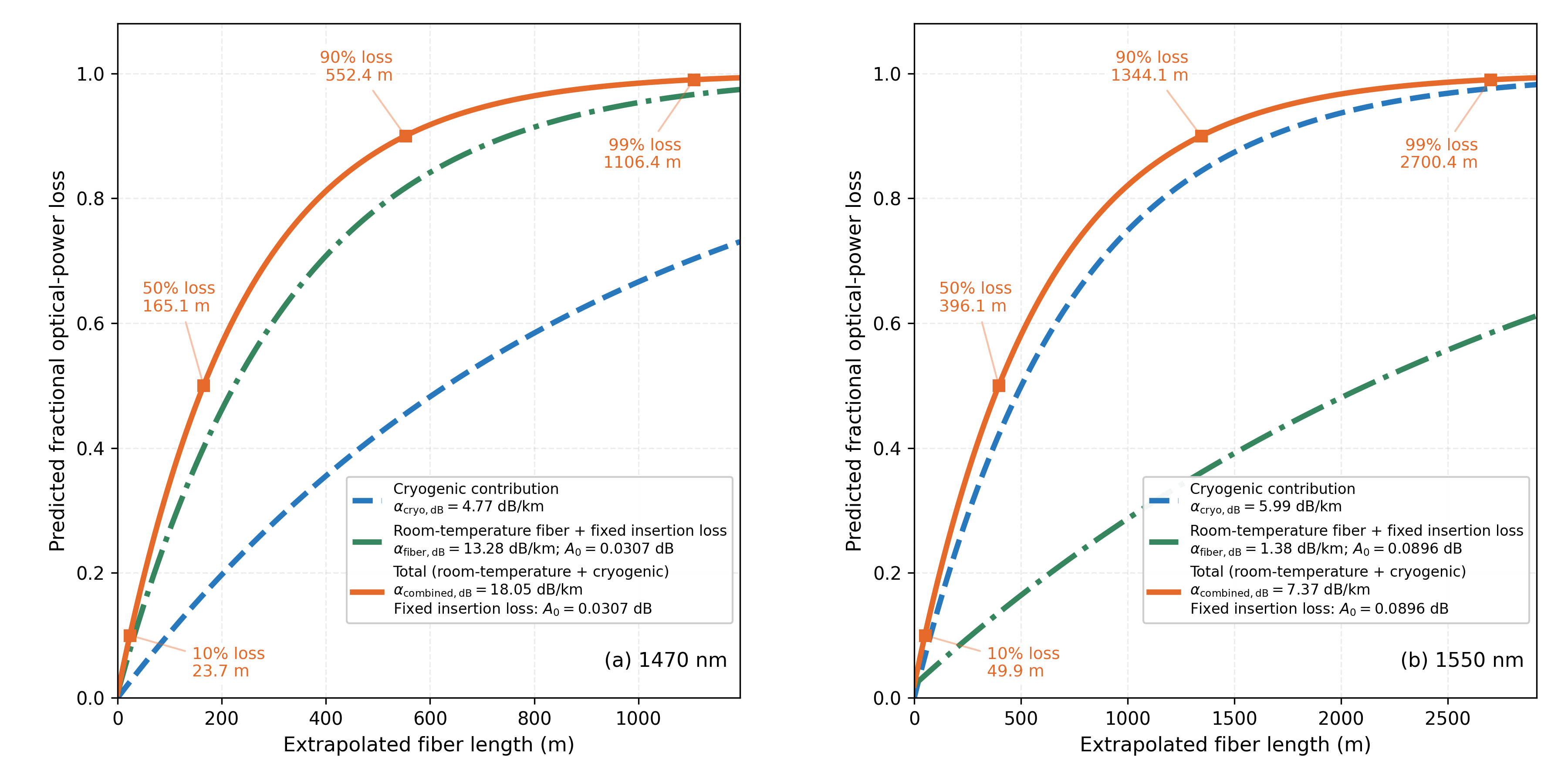}
    \caption{Projected fractional optical-power loss as a function of fiber length at (a) 1470~nm and (b) 1550~nm for the fiber studied in this work. The blue curves represent the induced cryogenic contribution, the green curves include room-temperature fiber attenuation and fixed insertion loss, and the orange curves represent the combined room-temperature and cryogenic losses.}
    \label{fig:effective_loss_projection}
\end{figure}

The attenuation parameters summarized in Table \ref{tab:attenuation_coefficients} were used to estimate the optical power losses as a function of fiber length for both wavelengths. At 1470~nm, the projected fiber lengths corresponding to total optical-power losses of 10\%, 50\%, 90\%, and 99\% are approximately 23.7, 165.1, 552.4, and 1106.4~m, respectively. At 1550~nm, the corresponding lengths are 49.9, 396.1, 1344.1, and 2700.4~m, respectively. Figure~\ref{fig:effective_loss_projection} shows the projected fractional optical-power loss as a function of fiber length at both wavelengths, allowing their optical transmission performance to be compared.

The calculated losses provide a useful comparison of the optical transmission performance at 1470 and 1550~nm and offer valuable guidance for selecting suitable fiber lengths in future PoF systems operating at cryogenic temperatures. 
\section{InGaAs optical power converter}
\label{sec:OPC}

In a PoF system, the OPC converts the optical power transmitted by the laser through the optical fiber into electrical power. These OPCs are also commonly referred to as photovoltaic power converters or laser power converters \cite{Mukherjee2013,Algora2022,Fafard2021Perspective}. Since OPCs convert optical power into electrical power through photon absorption across the semiconductor bandgap, without relying on moving parts, they could offer advantages for stable and reliable long-term power delivery to electronic systems.

\begin{table}[h!]
  \centering
  \caption{Main specifications of the commercially available Broadcom AFBR-POC205A8 and AFBR-POC205A9 OPCs ~\cite{Broadcom_POC205}. Typical values are taken from the manufacturer data sheet under the specified room-temperature operating conditions.}
  \vspace{1em}
  \begin{tabular}{lp{6.5cm}}
    Parameter & AFBR-POC205A8/AFBR-POC205A9 \\
    \hline
    Optimal input wavelength range (A8 / A9) & 1460-1500 / 1530-1570~nm \\
    Maximum optical input power & 1.5~W \\
    $P_\mathrm{mpp}$ at $P_\mathrm{in}=1.0$ / 1.5~W (typ.) & 0.45 / 0.69~W \\
    $V_\mathrm{mpp}$ (typ.) & 4.7~V \\
    $I_\mathrm{mpp}$ at $P_\mathrm{in}=1.0$ / 1.5~W (typ.) & 0.09 / 0.14~A \\
    Optimal load $R_\mathrm{mpp}$ (typ., 20~$^\circ$C) & $\approx 50.8~\Omega\cdot\mathrm{W}/P_\mathrm{in}$ \\
    Conversion efficiency & $>40\%$ \\
    Operating case temperature & $-40$ to $+85~^\circ$C \\
  \end{tabular}
  \label{tab:opc_specs}
\end{table}

The InGaAs OPCs characterized in this work are the Broadcom AFBR-POC205A8LT and AFBR-POC205A9  (see Table \ref{tab:opc_specs}), selected for operation at 1470 and 1550 nm, respectively. Both devices are multijunction optical power converters based on the vertical epitaxial heterostructure architecture (VEHSA) \cite{Broadcom_POC205}. While the AFBR-POC205A8 and AFBR-POC205A9 are commercially available OPCs designed for standard industrial applications operating temperatures (-40 - 85 $^{\circ}\text{C}$), the AFBR-POC205A8LT used in this work is a customized variant manufactured by Broadcom optmized for cryogenic operation at approximately 1470 nm \cite{Fafard2024}. The performance of an OPC depends on the match between the wavelength of the incident optical power and the spectral response of the semiconductor structure. The semiconductor bandgap determines the minimum photon energy required for absorption. Photons with energies below the bandgap cannot contribute to the corresponding band-to-band absorption, while for photon energies above the bandgap, part of the excess energy is lost through thermalization. Therefore, the semiconductor material and the wavelength of the incident optical power need to be properly matched for an efficient optical-to-electrical power conversion. This represents an important difference with respect to the PoF systems previously developed for DUNE prototypes, which operated at 808~nm using GaAs-based OPCs  \cite{arroyave2024characterization}. In the present work, the use of InGaAs-based OPCs allows the PoF system to operate at longer wavelengths such as 1470 and 1550 nm.

The wavelength response of the OPC also could be affected by the temperature, since the semiconductor bandgap changes with temperature. As the temperature decreases, the bandgap increases, resulting in a shift of the absorption edge toward shorter wavelengths \cite{Fafard2022_808}. This effect is particularly relevant for the present application, where the OPC is intended to operate at cryogenic temperature. 

In this section, we present the results of the characterization of the 1470 nm and 1550 nm OPCs at LN$_{2}$ temperature using different input powers on the laser transmitter.

\subsection{OPC characterization at cryogenic temperatures}

\begin{figure} [h!]
    \centering
    \includegraphics[width=\linewidth]{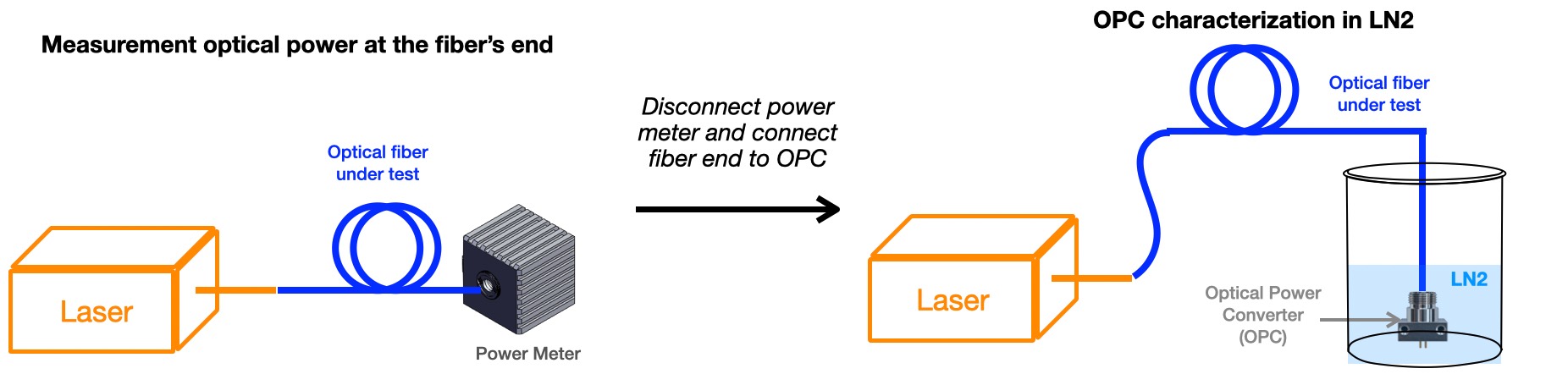}
    \caption{Schematic of the experimental setup used for the OPC characterization in LN$_2$. The optical power at the fiber's end was first measured using a power meter \cite{Thorlabs_S145C}. The power meter was then disconnected, and the same fiber's end was connected to the OPC and submerged in LN$_2$.}
    \label{fig:OPCTestStand}
\end{figure}

The test stand used for the OPC characterization is shown in Figure \ref{fig:OPCTestStand}. The AFBR-POC205A8LT and AFBR-POC205A9 OPCs were characterized at operating wavelengths of 1470 and 1550~nm, respectively, at LN$_2$ temperatures using various optical input powers delivered to each OPC.

\begin{figure}[h!]
    \centering
    \includegraphics[width=\linewidth]{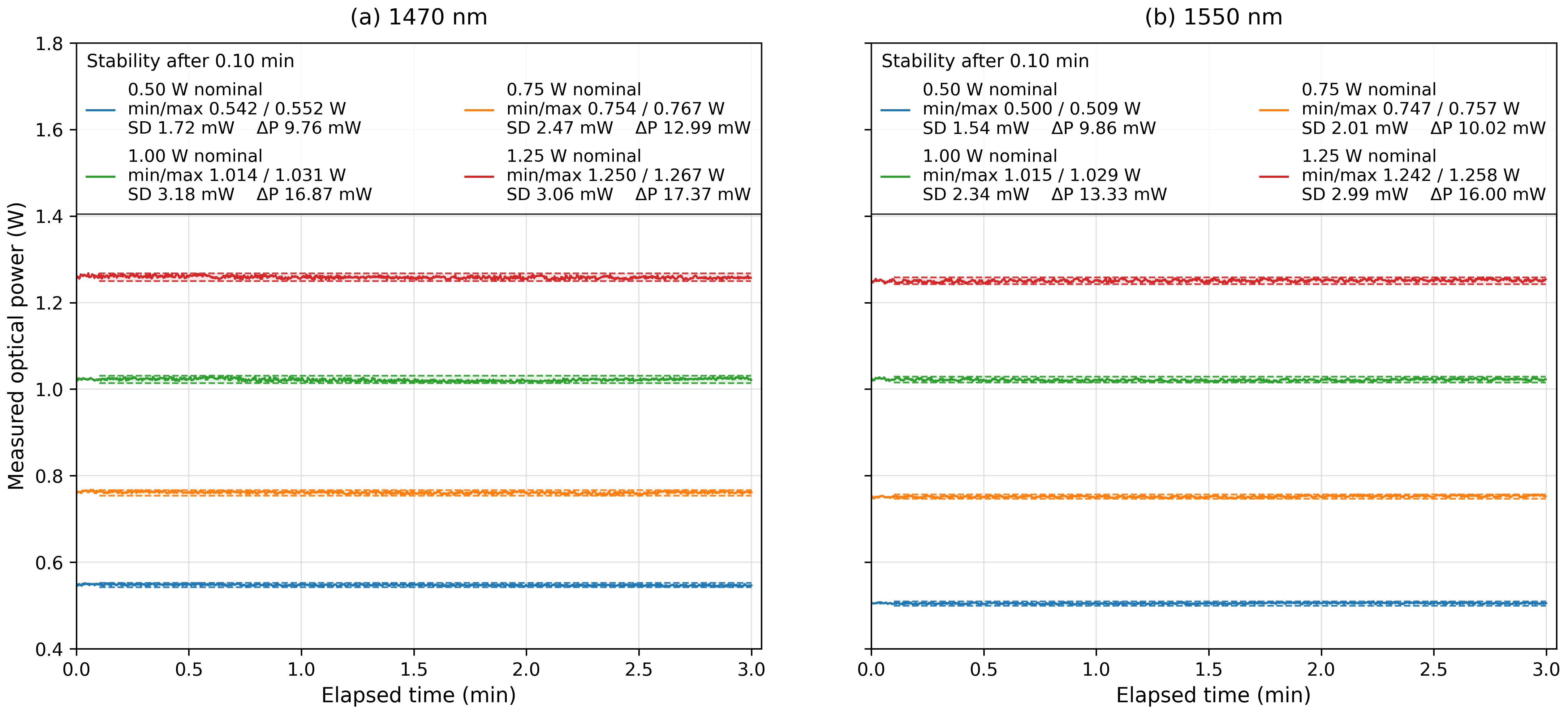}
    \caption{Measured optical power at the fiber output for the 1470 and 1550~nm lasers at nominal power settings of 0.50, 0.75, 1.00, and 1.25~W. Each trace shows a 3~min reference measurement used to determine the corresponding input power, $P_\mathrm{in}$. The minimum and maximum power, standard deviation, and peak-to-peak variation are reported for each configuration.}
    \label{fig:powerin}
\end{figure}

The optical power from the laser was delivered to the OPC through the GoPower optical fiber described in Section \ref{sec:opticalfibers}. Prior to the OPC characterization, the optical power delivered at the end of the fiber was measured using an integrating-sphere photodiode power meter. The laser transmitter power was adjusted to obtain nominal optical powers of 0.50, 0.75, 1.00, and 1.25 W. For each nominal power, the optical power was monitored for 3~min, and the mean measured value was used as the input power to the OPC, $P_\mathrm{in}$. The power stability was characterized by the standard deviation and the minimum-to-maximum variation, as shown in Figure \ref{fig:powerin}. 
This procedure was applied for the measurements obtained with the 1470 nm and 1550 nm laser transmitters.

After the optical input power was measured, the fiber was connected to the corresponding OPC. For the cryogenic measurements, the OPC was fully immersed in LN$_2$ at approximately $77~\mathrm{K}$ ($-196\,^\circ\mathrm{C}$) in a Dewar (see Figure \ref{fig:OPCTestStand}). Only a few centimeters of optical fiber needed to reach the OPC were immersed, while the remaining fiber was kept outside the Dewar. This arrangement minimized the additional losses associated with cooling and thermal contraction, as discussed in Section \ref{sec:opticalfibers}. Measurements started 10~min after immersion to allow the OPC to reach thermal equilibrium and the laser output to stabilize. The electrical response of the OPC was measured using a Source Meter Unit (SMU) connected in a four-wire (Kelvin) configuration. For each optical input power, the OPC terminal voltage was swept from 0~V, corresponding to the short-circuit condition, to a voltage near the open-circuit voltage, $V_\mathrm{oc}$. The current was measured at each voltage step to obtain the corresponding current-voltage (I-V) curve. The I-V measurement was repeated three times for each OPC and optical input power.

\subsubsection{Current-voltage characterization}

\begin{figure}[h!]
    \centering
    \includegraphics[width=1.0\linewidth]{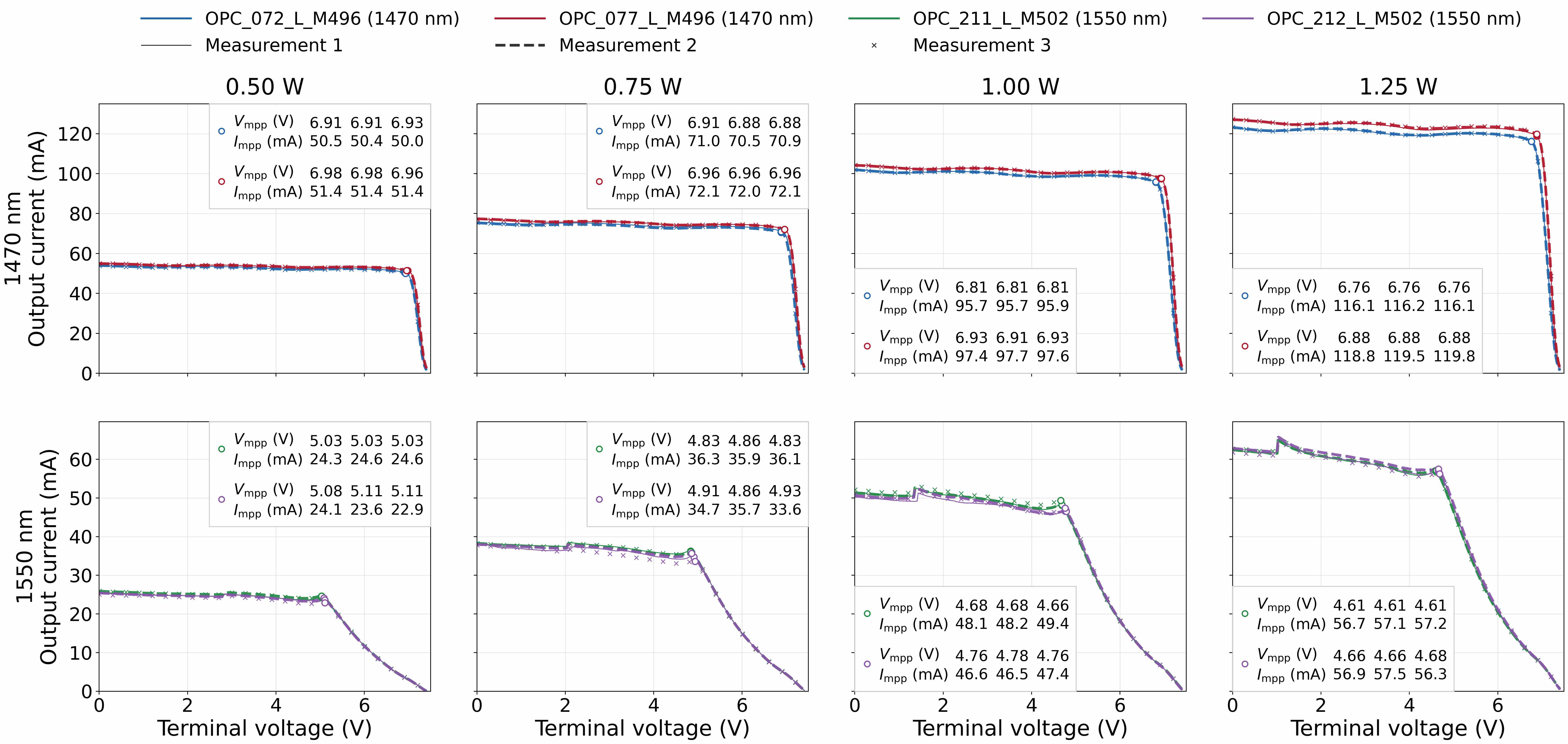}
    \caption{Current-voltage characteristics of the OPCs measured in LN$_2$ for optical input powers from 0.50 to 1.25~W. The upper row corresponds the 1470~nm OPCs and the lower row the 1550~nm OPCs. Three repeated sweeps are shown for each device and input power. The legends report the $V_{\mathrm{mpp}}$ and $I_{\mathrm{mpp}}$ obtained from each of the three sweeps. Open markers indicate the MPP identified from each I-V sweep.}
    \label{fig:ivpower}
\end{figure}

Figure~\ref{fig:ivpower} shows the I-V curves measured in LN$_2$ for
the 1470 and 1550~nm OPCs. At 1470~nm, the output current remains approximately constant over most of the voltage range and then decreases sharply as the voltage approaches $V_\mathrm{oc}$. As $P_\mathrm{in}$ increases, the output current increases from approximately 50~mA at the lowest input power to about 120~mA at the highest input power. The voltage at the maximum power point (MPP) remains close to 6.8-7.0~V over the measured range. The three repeated measurements for each OPC show good agreement.

The 1550~nm OPCs show a different response. For the same optical input power, the output current is lower and the decrease in current near the open-circuit region is more gradual. The voltage at the MPP decreases from approximately 5.0~V at the lowest input power to about 4.6~V at the highest input power, while the corresponding current increases from approximately 24 mA to 57~mA. 

\subsubsection{Electrical output power and optimum load}

The measured I-V curves were used to calculate the OPC electrical output
power ($P_\mathrm{out}$) and load resistance at each measurement point as:

\begin{equation}
    R = \frac{V}{|I|}, \qquad P_\mathrm{out} = V|I|.
\end{equation}

\begin{figure}[h!]
    \centering
    \includegraphics[width=1\linewidth]{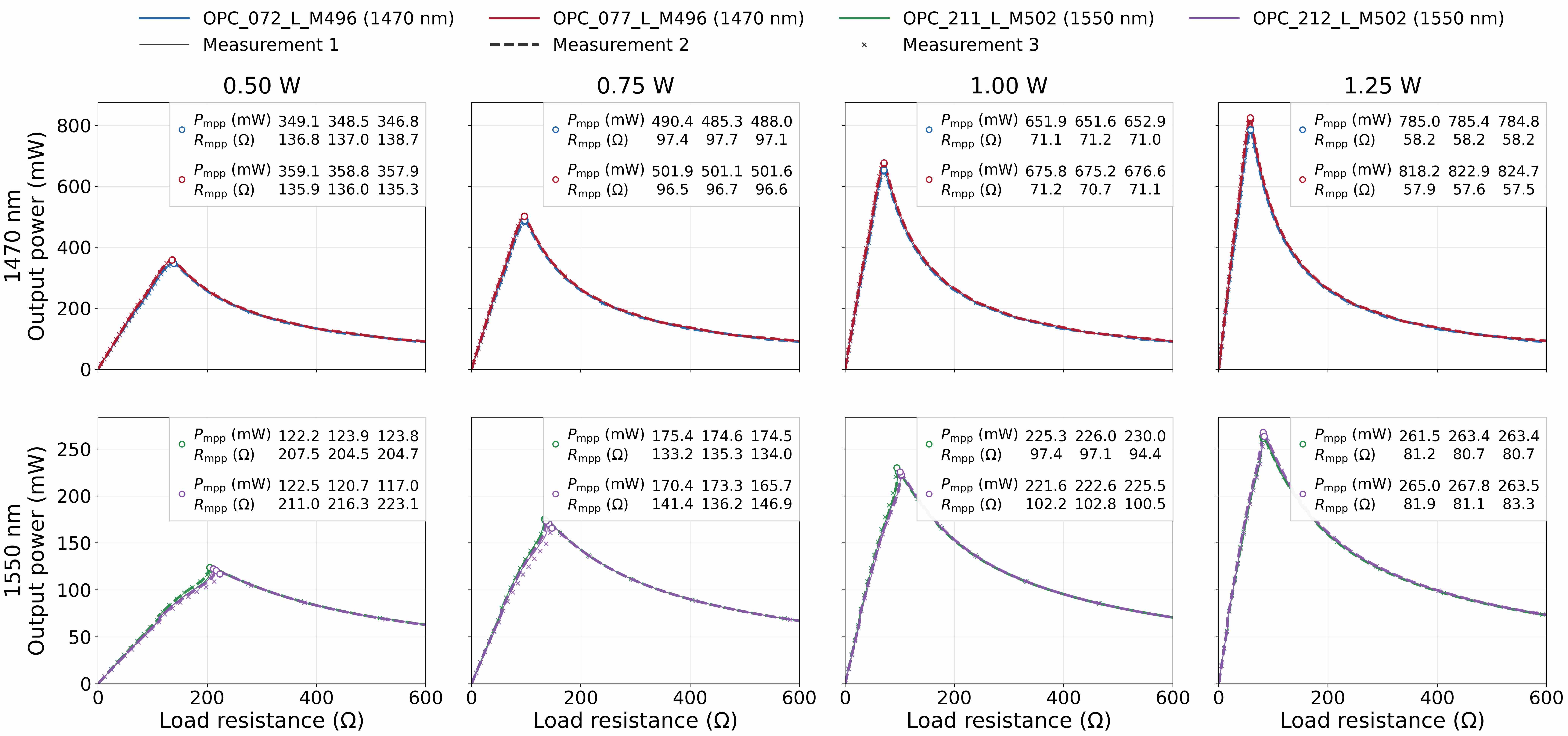}
    \caption{Electrical output power as a function of calculated load resistance for the OPCs measured in LN$_2$. The upper row corresponds to 1470~nm and the lower row to 1550~nm. Three repeated measurements are shown for each OPC and input power. The legends report the $P_{\mathrm{mpp}}$ and $R_{\mathrm{mpp}}$ obtained from each of the three sweeps. Open markers indicate the maximum power point and the corresponding optimum load.}
    \label{fig:powerload}
\end{figure}

Figure~\ref{fig:powerload} shows the resulting OPC electrical output power as a function of the calculated load resistance. For each input power, the OPC output power increases with load resistance until reaching a clear maximum and then decreases as the OPC approaches the open-circuit region. The resistance at this maximum therefore defines the optimum load, $R_\mathrm{mpp}$, for the corresponding operating condition.

For the 1470~nm OPCs, $R_\mathrm{mpp}$ decreases from approximately 137~$\Omega$ at the lowest input power to  $\sim$ 58~$\Omega$ at the highest while the maximum OPC electrical output power increases from approximately 0.35 to 0.82~W. Similarly, for the 1550~nm OPCs, $R_\mathrm{mpp}$ decreases from 205-220~$\Omega$ to approximately 81~$\Omega$ across the same input power range, while the maximum output power increases from $\sim$0.12 W to 0.27 W. Thus, for both wavelengths, increasing the optical input power shifts the maximum power point toward lower load resistance.

\begin{figure}[h!]
    \centering
    \includegraphics[width=1\linewidth]{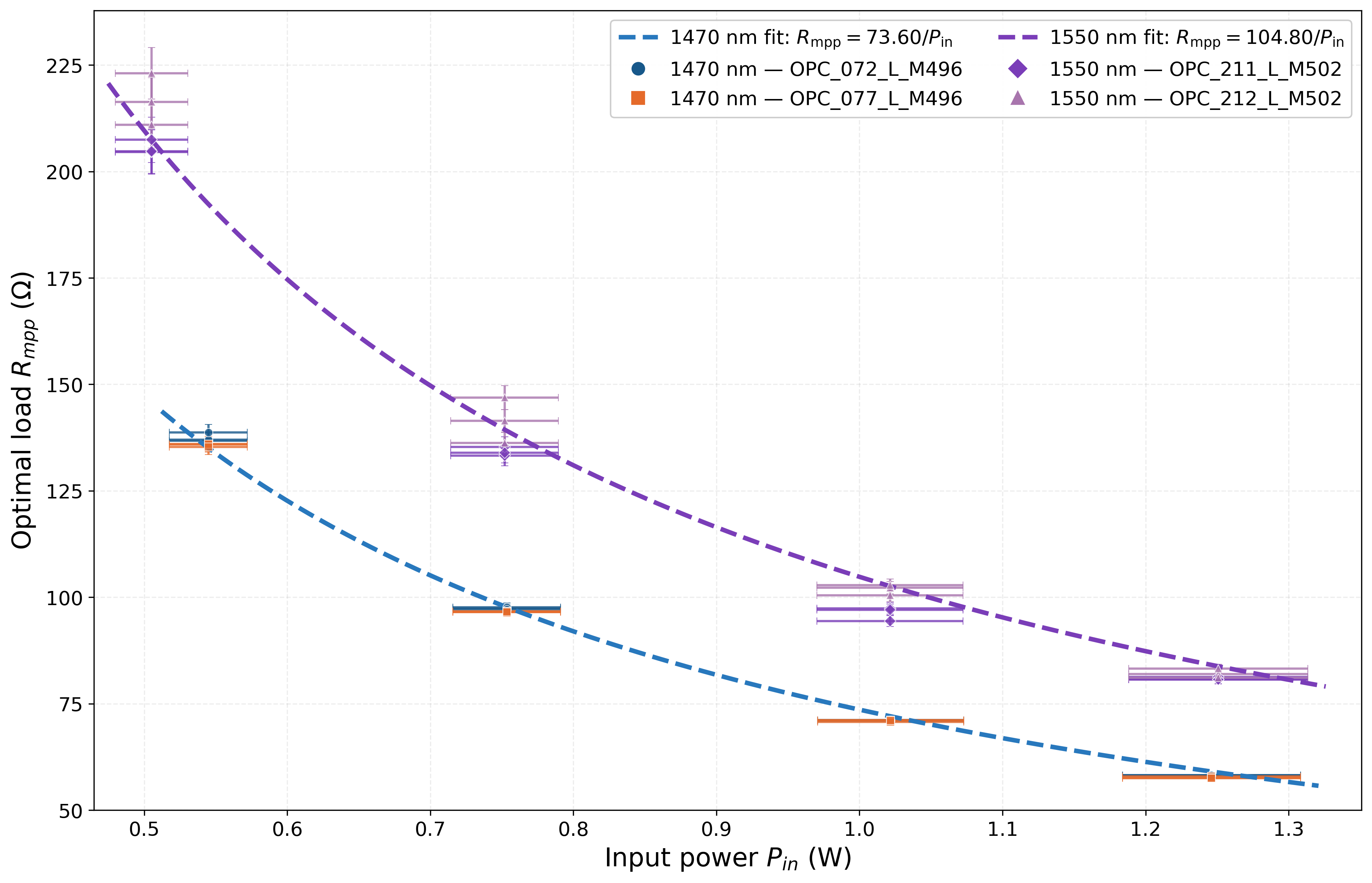}
    \caption{Optimum load resistance, $R_\mathrm{mpp}$, as a function of the measured optical input power for the OPCs operated in LN$_2$ at 1470 and 1550~nm. Individual points correspond to the values obtained from each I-V sweep for the two OPCs tested at each wavelength. The dashed curves show fits of the form $R_\mathrm{mpp}=A/P_\mathrm{in}$.}
    \label{fig:rmpp_pin}
\end{figure}

The dependence of the optimum load on the OPC optical input power is shown in Figure \ref{fig:rmpp_pin}. For both wavelengths, $R_\mathrm{mpp}$ decreases as $P_\mathrm{in}$ increases and is well described by an inverse dependence on the OPC input power.

\subsubsection{Optical-to-electrical power conversion efficiency}

For each I-V sweep, the maximum value of $P_\mathrm{out}$ was identified as the maximum power point, $P_\mathrm{mpp}$. The corresponding values of $V_\mathrm{mpp}$, $I_\mathrm{mpp}$, and $R_\mathrm{mpp}$ were also recorded. The optical-to-electrical power conversion efficiency was calculated as:

\begin{equation}
    \eta = \frac{P_\mathrm{mpp}}{P_\mathrm{in}}.
\end{equation}

\begin{figure}[!h]
    \centering
    \includegraphics[width=1\linewidth]{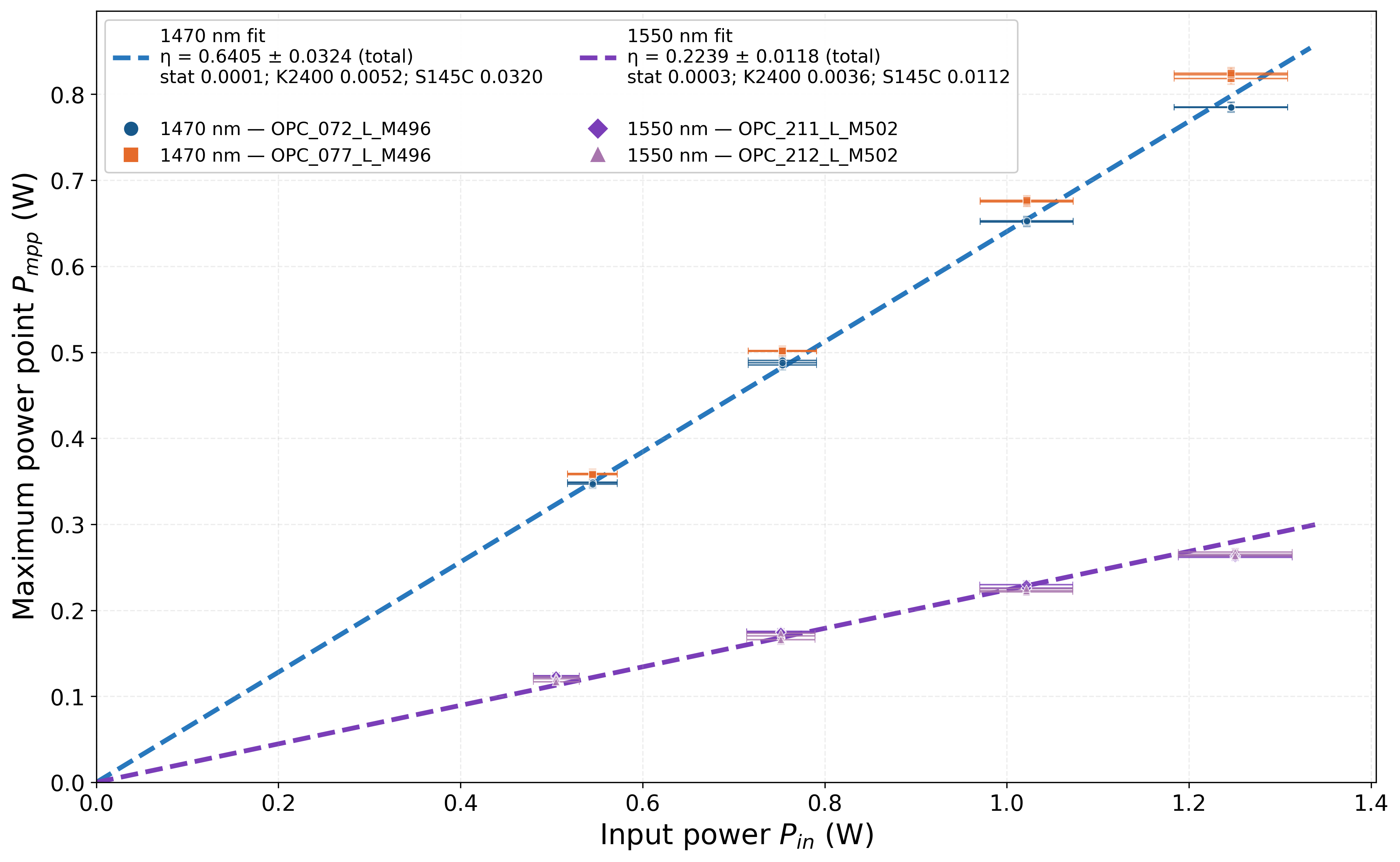}
    \caption{Maximum electrical output power, $P_\mathrm{mpp}$, as a function of the measured optical input power for OPCs operated in LN$_2$. Individual points correspond to the repeated I--V sweeps for the two OPCs tested at each wavelength. Dashed lines show linear fits constrained through the origin, with the fitted slope corresponding to the optical-to-electrical power conversion efficiency.}
    \label{fig:powereff}
\end{figure}

Figure~\ref{fig:powereff} shows $P_\mathrm{mpp}$ as a function of the measured optical input power $P_\mathrm{in}$ for all individual sweeps. The repeated measurements at each input-power setting were used as individual points and included in the linear fit, where the fitted slope, $\eta$, represents the overall optical-to-electrical power conversion efficiency over the measured power range.

\begin{equation}
    P_\mathrm{mpp} = \eta P_\mathrm{in},
\end{equation}

For the 1470~nm OPCs, the maximum output power follows an approximately linear dependence on the optical input power, giving a conversion efficiency of $\eta = 0.6405 \pm 0.0324$. The 1550~nm OPCs show a substantially lower maximum output power, with a fitted efficiency of $\eta = 0.2239 \pm 0.0118$. The lower efficiency measured for the AFBR-POC205A9 OPCs at cryogenic temperatures is consistent with their design for standard operating temperatures (-40 to 85 $^\circ\text{C}$), rather than cryogenic applications. Under the tested conditions, the AFBR-POC205A8LT operating at 1470 nm showed better performance, making it the preferred option for cryogenic operation. The 1470~nm devices converted approximately 64\% of the measured optical input power into electrical output power at LN$_{2}$ temperature test conditions, compared with approximately 22\% for the commercial 1550~nm devices. The characterized GaAs OPCs at cryogenic conditions presented an optical-to-electrical efficiency between 50 to 55 \% \cite{arroyave2024characterization}. The results found for the characterization of InGaAs operating at 1470 nm presented an increase of about $10$ percentage points in conversion efficiency compared with the GaAs OPCs when operating in cryogenic temperatures.

The uncertainty reported for $\eta$ also includes the systematic contributions from the optical- and electrical-power measurements. The fitted efficiency $\eta$ and its statistical uncertainty, $u_{\mathrm{stat}}(\eta)$, were obtained directly from the ROOT minimization of $P_{\mathrm{mpp},i} - \eta P_{\mathrm{in},i}$. The remaining contributions were treated as shared systematic uncertainties. The $\pm5\%$ accuracy of the integrating-sphere photodiode power sensor S145C gives $u_{\mathrm{opt}}(\eta)=0.05|\eta|$. For the Keithley 2400 SMU, the voltage and current uncertainties were taken from the instrument specifications,

\begin{equation}
u_{V,i}=1.5\times10^{-4}|V_i|+1.5~\mathrm{mV},
\qquad
u_{I,i}=2.2\times10^{-3}|I_i|+570~\mu\mathrm{A},
\end{equation}
and propagated through $P_{\mathrm{mpp}}=VI$ to the fitted slope using its sensitivity to each measurement,

\begin{equation}
u_{\mathrm{smu}}^2(\eta)=
\left(\sum_i c_i I_i u_{V,i}\right)^2+
\left(\sum_i c_i V_i u_{I,i}\right)^2,
\qquad
c_i=\frac{w_iP_{\mathrm{in},i}}
{\sum_j w_jP_{\mathrm{in},j}^2},
\qquad
w_i=\frac{1}{s_{g(i)}^2}.
\end{equation}

The coefficient $c_i$ represents the sensitivity of the fitted efficiency to the output power of point $i$, while $w_i$ is its statistical weight. Here, $P_{\mathrm{in},i}$ is the optical input power for measurement $i$, $g(i)$ identifies the group of three measurements taken with a given OPC at a given optical-power setting, and $s_{g(i)}$ is the standard deviation of the corresponding $P_{\mathrm{mpp}}$ values.
    
For the uncertainty propagation, the Keithley 2400 SMU voltage and current uncertainties were treated as common systematic contributions across the repeated sweeps. The voltage and current contributions were propagated separately and assumed to be independent of each other.

The reported uncertainty on the efficiency was then obtained by combining the statistical and independent systematic contributions in quadrature,

\begin{equation}
u(\eta)=
\sqrt{
u_{\mathrm{stat}}^2(\eta)
+u_{\mathrm{opt}}^2(\eta)
+u_{\mathrm{smu}}^2(\eta)
}.
\end{equation}
\section{Light leakage studies for PoF systems operating at 1470 nm and 1550 nm wavelengths}
\label{sec:light_leak}

The light leakage test stands were designed to characterize the response of silicon photomultipliers (SiPMs) to IR light escaping from the optical fibers and OPCs used in the PoF systems. The light leakage measurements were performed using a 50~L Dewar that accommodated the components under test and a string of twenty Broadcom SiPMs, each with a $6 \times$ 6 mm$^2$ active area, mounted on a flexible printed-circuit board \cite{BroadcomSiPM2025}, referred to as the SiPM flex in this paper. This section presents the experimental setups and data analysis procedures used to obtain the light-leakage measurements at different IR wavelengths.

\subsection{Setup}

\begin{figure}[h!]
    \centering
    \includegraphics[width=\linewidth]{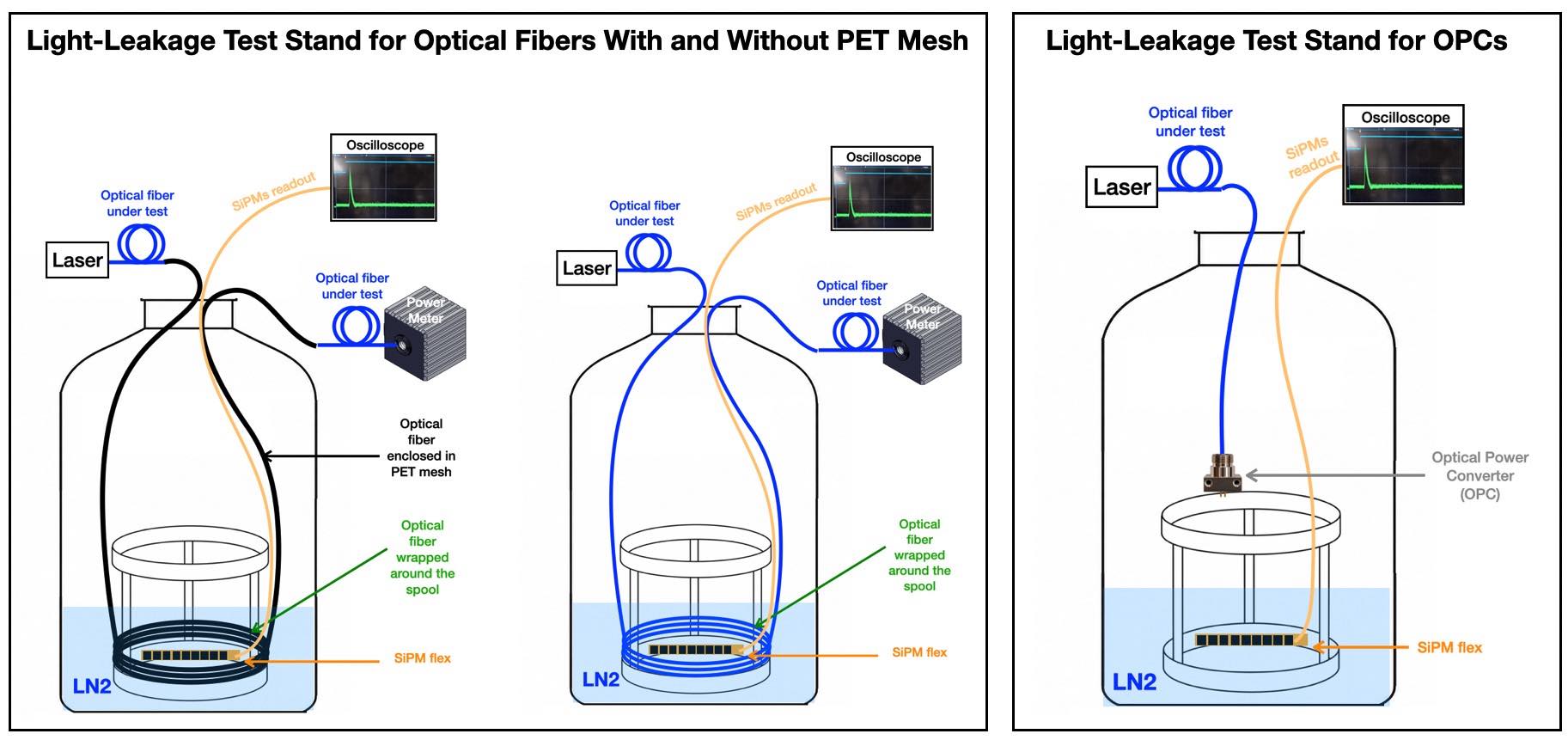}
    \caption{Schematic of the experimental setups used for the light-leakage measurements. Left: test stand used to measure light leakage from fibers under test with and without the PET mesh, with the fiber wrapped around the spool above the SiPM flex. Right: test stand used to measure light leakage from the OPC, with the OPC and a small portion of the connected optical fiber inserted into the Dewar. The SiPM signals were read out directly with an oscilloscope.}
    \label{fig:LightTestStand}
\end{figure}

Two test stands were used for the light leakage measurements, as shown in Figure \ref{fig:LightTestStand}. The first test stand was used to measure the light leakage from the optical fiber with and without the PET mesh. A spool holder was built to maintain the fiber configuration and the relative position between the fiber under test and the SiPM flex throughout all the measurements. The laser transmitter was connected directly to one end of the fiber under test, while the other end was connected to power meter located outside the Dewar, as shown in Figure \ref{fig:LightTestStand} (left). The optical fiber under test was wrapped around a spool with a radius of 8 cm, with the SiPM flex positioned below the fiber loops. The active areas of the SiPMs were oriented toward the interior of the loops to increase the acceptance for light emitted from the fiber under test. The same fiber-to-SiPM flex configuration was maintained throughout the measurements. The second test stand was used to measure the light leakage contribution from the OPC, rather than from the tested fiber. In this configuration, the output end of the fiber was connected to the OPC. The OPC and about 10 cm of the connected fiber were inserted into the Dewar, while the remaining fiber was kept outside, as shown in Figure \ref{fig:LightTestStand} (right).

For all measurements, the signals from the SiPM flex were amplified using dedicated readout electronics. The amplified electrical signals from the SiPM flex were read out directly with a Tektronix MDO4104C oscilloscope. Waveforms were acquired and stored using custom Python scripts developed to automate the data acquisition. The measurements were performed using laser transmitters operating at 808, 1470, and 1550~nm. The 808~nm wavelength provided a reference for the light-leakage measurements, as the SiPMs are sensitive to incident photons at this wavelength. The use of 1470 and 1550 nm was pursued aimed at reducing the impact of optical light leakage on the SiPM flex response. The comparison among the three wavelengths was therefore used to evaluate the impact of PoF light leakage on the SiPM flex response.

The light-leakage measurements were performed using the same type of optical fiber (see Table \ref{tab:fiber_specs}) with two different fiber jacket configurations: light-tight and non-light-tight. The measurements were performed with and without the PET mesh at 808, 1470, and 1550 nm. The OPC measurements were performed at 1470 and 1550 nm. Tables \ref{tab:fiber_light_leakage} and \ref{tab:opc_light_leakage} summarize the configurations used for the fiber and OPC light-leakage measurements, respectively.

\begin{table}[!h]
    \centering
    \caption{Summary of the configurations used for the fiber light-leakage measurements. Both fibers are of the same type but differ in the light-blocking capabilities of their jackets. The length on the spool corresponds to the portion of the optical fiber wrapped around the spool during the measurements.}
    \vspace{1em}
    \setlength{\tabcolsep}{4pt}
    \begin{tabular}{@{}lcccc@{}}
    Fiber & Configuration & Length on spool & $\lambda$ (nm) & Laser Power (mW) \\
    \hline
    \multirow{3}{*}{Light-tight fiber}
    & \multirow{3}{*}{1 mesh}
    & \multirow{3}{*}{20 m}
    & 808  & 1000 \\
    & & & 1470 & 1200 \\
    & & & 1550 & 1200 \\
    \hline
    \multirow{3}{*}{Non-light-tight fiber}
    & \multirow{3}{*}{1 mesh}
    & \multirow{3}{*}{20 m}
    & 808  & 1000 \\
    & & & 1470 & 1200 \\
    & & & 1550 & 1200 \\
    \hline
    \multirow{2}{*}{Non-light-tight fiber}
    & \multirow{2}{*}{No mesh}
    & \multirow{2}{*}{8 m}
    & 1470 & 1200 \\
    & & & 1550 & 1200 \\
    \end{tabular}\label{tab:fiber_light_leakage}
\end{table}

\begin{table}[h!]
    \centering
    \caption{Summary of the configurations used for the OPC light-leakage measurements at 1470 and 1550 nm.}
    \vspace{1em}
    \begin{tabular}{lccc}
    Fiber & $\lambda$ (nm) & Laser Power (mW) & OPC \\
    \hline
    \multirow{2}{*}{Light-tight fiber}
    & 1470 & 1200 & AFBR-POC205A8L\\
    & 1550 & 1200 & AFBR-POC205A9\\
    \hline
    \multirow{2}{*}{Non-light-tight fiber}
    & 1470 & 1200 & AFBR-POC205A8L\\
    & 1550 & 1200 & AFBR-POC205A9\\
    \end{tabular}
    \label{tab:opc_light_leakage}
\end{table}

\subsection{Analysis}
\label{sed:LKAnalysis}
The described processing chain was applied to both the dark-count rate (DCR) and light leak measurements acquired with the PoF system presented in the following subsections. The analysis used the recorded SiPM flex waveforms and consisted of the following steps:

\begin{itemize}
    \setlength{\itemsep}{0pt}
    \setlength{\parskip}{0pt}
    \setlength{\parsep}{0pt}
    \item [\textbf{1.}] Waveform preparation
    \item [\textbf{2.}] Single-photoelectron response template
    \item [\textbf{3.}] Noise template
    \item [\textbf{4.}] Deconvolution and filtering
    \item [\textbf{5.}] Peak identification
\end{itemize}

\subsubsection{Waveform preparation}

The waveforms acquired with the oscilloscope were initially stored as CSV files to later be saved as histograms in ROOT files. Each waveform was stored together with its metadata, including the sampling interval, record length, and voltage units. The DC offset was calculated independently for each waveform. Let $x_i[n]$ denote the measured voltage at sample $n$ of waveform $i$. The baseline of waveform $i$, denoted by $B_i$, was calculated as:

\begin{equation}
B_i =
\frac{1}{N_B}
\sum_{n=n_{\mathrm{first}}}^{n_{\mathrm{last}}}
x_i[n],
\end{equation}

where $n_{\mathrm{first}}$ and $n_{\mathrm{last}}$ define the selected baseline interval and $N_B$ is the number of samples contained in this interval. The baseline-subtracted waveform $x_i^{\mathrm{corr}}[n]$ was defined as

\begin{equation}
x_i^{\mathrm{corr}}[n] = x_i[n]-B_i.
\end{equation}

where $x_i^{\mathrm{corr}}[n]$ represents the corrected voltage at sample $n$ after removing the DC offset of waveform $i$. The waveforms acquired during the light-leakage measurements were organized according to the experimental configurations summarized in Tables \ref{tab:fiber_light_leakage} and \ref{tab:opc_light_leakage}. The same data-processing procedure was applied to all configurations to allow a consistent comparison of the SiPM flex response under the different measurement conditions.

\subsubsection{Single-photoelectron response template}

\begin{figure}[!ht]
    \centering
    \includegraphics[width=1.0\linewidth]{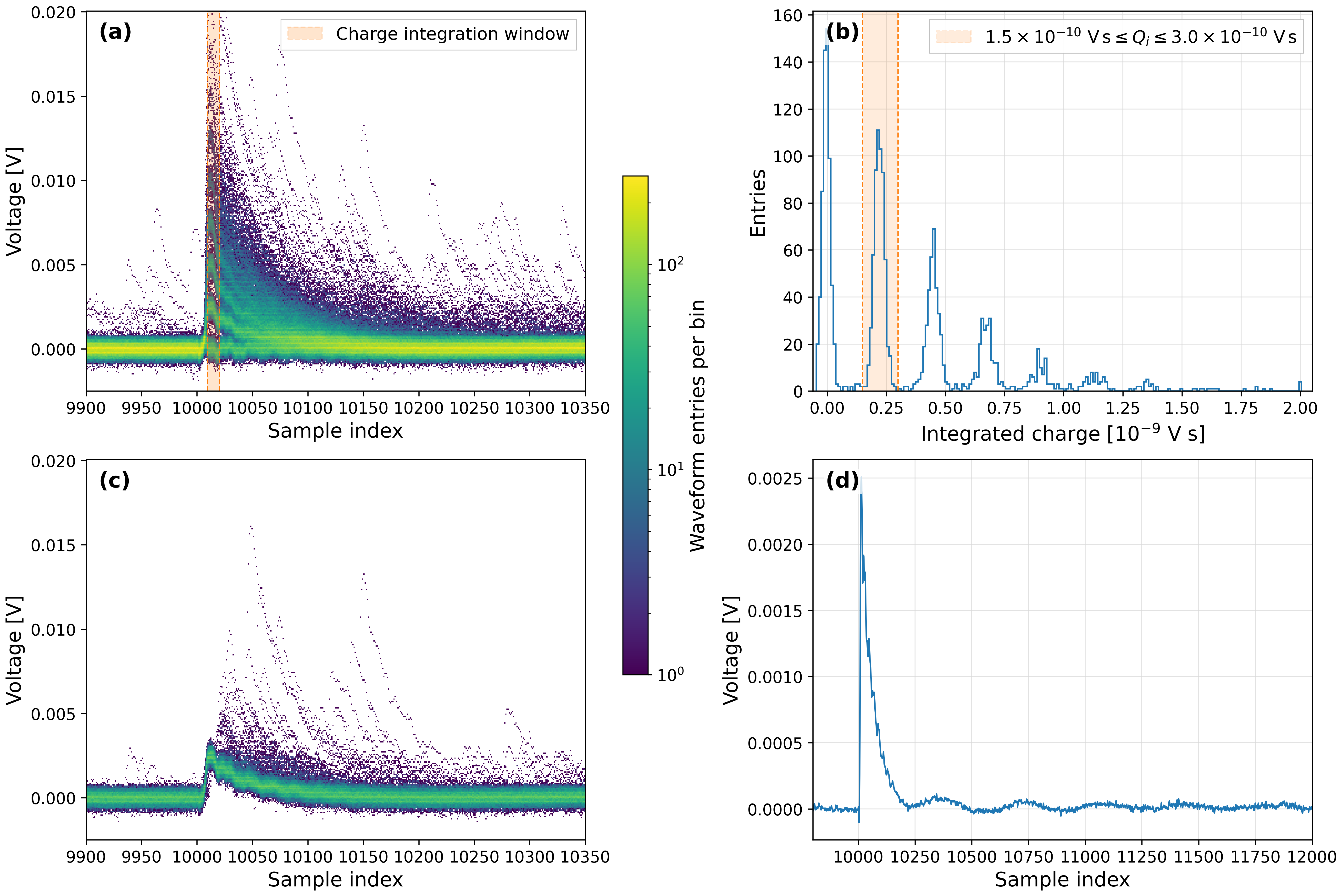}
    \caption{Construction of the single-photoelectron response template. (a) Persistence distribution of the LED waveforms, with the integration interval used to calculate the pulse area indicated by the shaded region. (b) Distribution of the integrated pulse area, with the single-p.e. selection indicated by the shaded interval. (c) Persistence distribution of the waveforms passing the single-p.e. selection. (d) Average response obtained from the selected waveforms and used as the single-p.e. response template for the deconvolution.}
    \label{fig:spe_template}
\end{figure}

A dedicated data set using a blue light LED was obtained to understand the single-photoelectron (single p.e.) response of the Broadcom SiPM flex submerged in liquid nitrogen. The persistence distribution was first inspected to identify the pulse position and define the integration interval, as shown in figure~\ref{fig:spe_template}(a). The integrated pulse area was calculated using the trapezoidal rule:

\begin{equation}
Q_i =
\Delta t
\sum_{n=n_a}^{n_b-1}
\frac{
x_i^{\mathrm{corr}}[n]+x_i^{\mathrm{corr}}[n+1]
}{2},
\end{equation}

where $\Delta t$ is the sampling interval and $n_a$ and $n_b$ define the lower and upper limits of the integration interval. Since the impedance and electronics gain are not included, $Q_i$ is reported in V\,s. The distribution of $Q_i$ shown in Figure~\ref{fig:spe_template}(b) was used to select the single p.e. population. Waveforms with $Q_i$ in the interval:

\begin{equation}
1.5\times10^{-10}\ {\rm V\,s}
\leq Q_i \leq
3.0\times10^{-10}\ {\rm V\,s}
\end{equation}

were selected, resulting in 596 waveforms out of the 2000 LED waveforms analyzed. The selected waveforms are shown in figure~\ref{fig:spe_template}(c). The single-p.e. response template was obtained by averaging the selected baseline-subtracted waveforms (see Eq. \ref{eq:5.5}), where $N_{\mathrm{sel}}$ is the number of waveforms passing the single-p.e. selection. The resulting response is shown in Figure~\ref{fig:spe_template}(d) and was used as the single-p.e-response template for the deconvolution.

\begin{equation}
r[n] =
\frac{1}{N_{\mathrm{sel}}}
\sum_{i\in\mathrm{single\mbox{-}p.e.}}
x_i^{\mathrm{corr}}[n].
\label{eq:5.5}
\end{equation}

\subsubsection{Noise template}

The noise template was constructed from a separate data set acquired while the SiPM flex bias was set below the breakdown voltage. No LED illumination was applied, and the PoF system was not operating during these measurements. The Fourier transform of each noise waveform was calculated using the FFT routines provided by ROOT. For a noise waveform $\eta_i[n]$, the transform can be written as:

\begin{equation}
\widetilde{\eta}_i[k]
=
\sum_{n=0}^{N-1}
\eta_i[n]\,
e^{-2\pi i kn/N}.
\end{equation}

For every FFT bin, the code calculated the spectral power directly from the real and imaginary components returned by ROOT:

\begin{equation}
\left|\widetilde{\eta}_i[k]\right|^2
=
\operatorname{Re}\!\left(\widetilde{\eta}_i[k]\right)^2
+
\operatorname{Im}\!\left(\widetilde{\eta}_i[k]\right)^2.
\end{equation}

The noise power template was obtained by averaging this quantity over all accepted noise waveforms:

\begin{equation}
P_N[k]
=
\frac{1}{N_{\mathrm{noise}}}
\sum_{i=1}^{N_{\mathrm{noise}}}
\left|\widetilde{\eta}_i[k]\right|^2.
\label{eq:noise}
\end{equation}

\begin{figure}[!h]
    \centering
    \includegraphics[width=\linewidth]{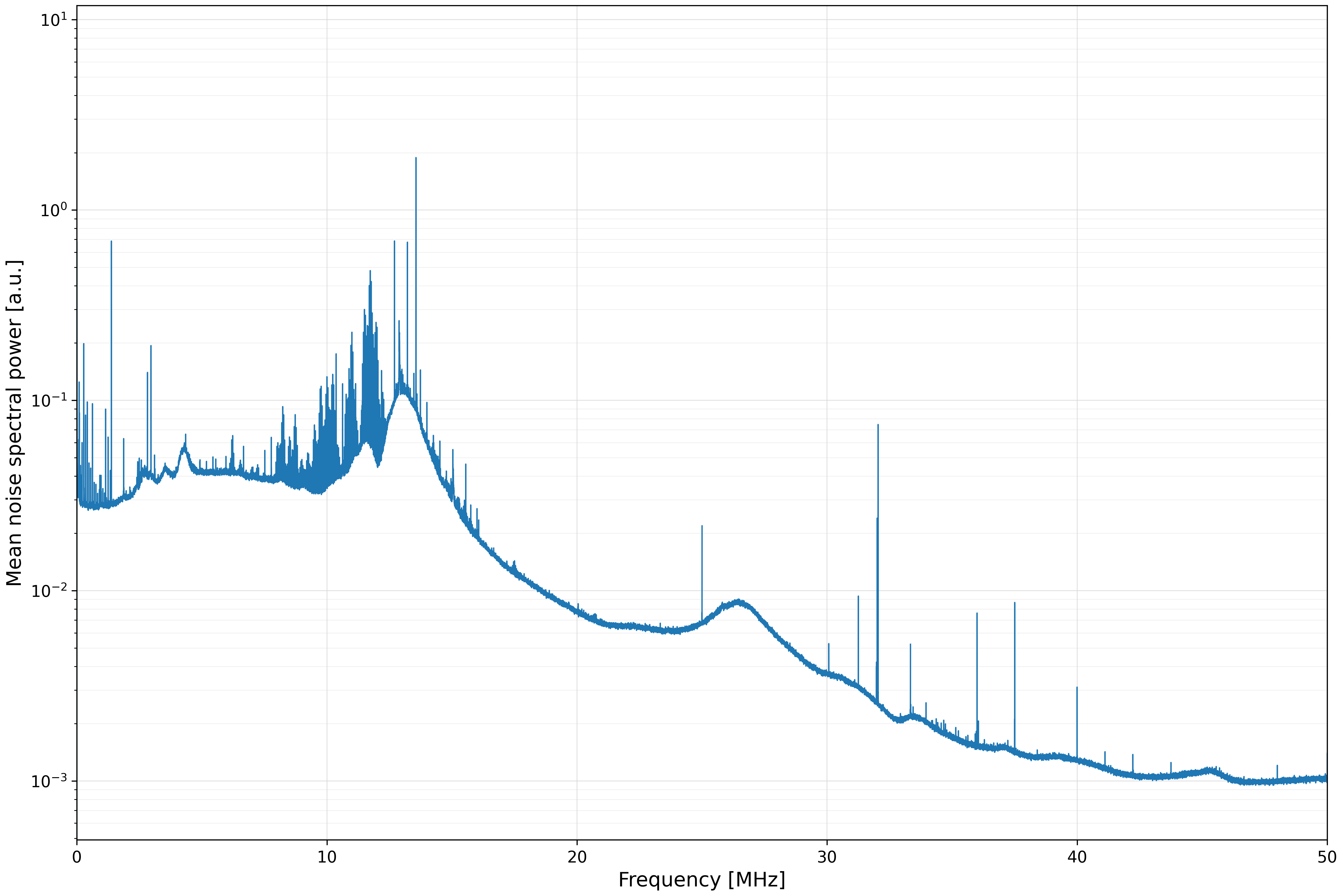}
    \caption{Average noise power spectrum obtained from the 5000 noise waveforms used to construct $P_N[k]$. The spectrum represents the frequency-dependent noise included in the Wiener deconvolution.}
    \label{fig:noise_spectrum}
\end{figure}

The noise template was constructed by averaging the power spectra of 5000 noise waveforms. The resulting spectrum, shown in Figure~\ref{fig:noise_spectrum}, characterizes the frequency-dependent noise of the SiPM flex and readout chain and was included in the Wiener filter to suppress noise-dominated frequency components during the Wiener deconvolution. 

\subsubsection{Deconvolution and filtering}

The measured waveform $x[n]$ was modeled as a sequence of photon-induced pulses convolved with the single-p.e.\ response of the SiPM flex and readout electronics:

\begin{equation}
x[n] = (p*r)[n] + \eta[n],
\end{equation}

where $p[n]$ is the underlying pulse sequence, $r[n]$ is the measured single-p.e.\ response, $\eta[n]$ represents the electronic noise of the system, and $*$ denotes convolution. In the absence of noise, the corresponding Fourier-domain relation is:

\begin{equation}
X[k] = P[k]R[k].
\end{equation}

Recovering the pulse sequence would therefore ideally require division by the single-p.e.\ response in the Fourier space, $P[k]=X[k]/R[k]$. This can also be understood by defining an inverse filter that transforms the single-p.e.\ response into a delta function:

\begin{equation}
(r*h_{\delta})[n]=\delta[n].
\end{equation}

Since the delta function is the identity for convolution and $\mathcal{F}\{\delta[n]\}=1$, the corresponding frequency-domain filter is:

\begin{equation}
H_{\delta}[k]=\frac{1}{R[k]}.
\end{equation}

\begin{figure}[!ht]
    \centering
    \includegraphics[width=1\linewidth]{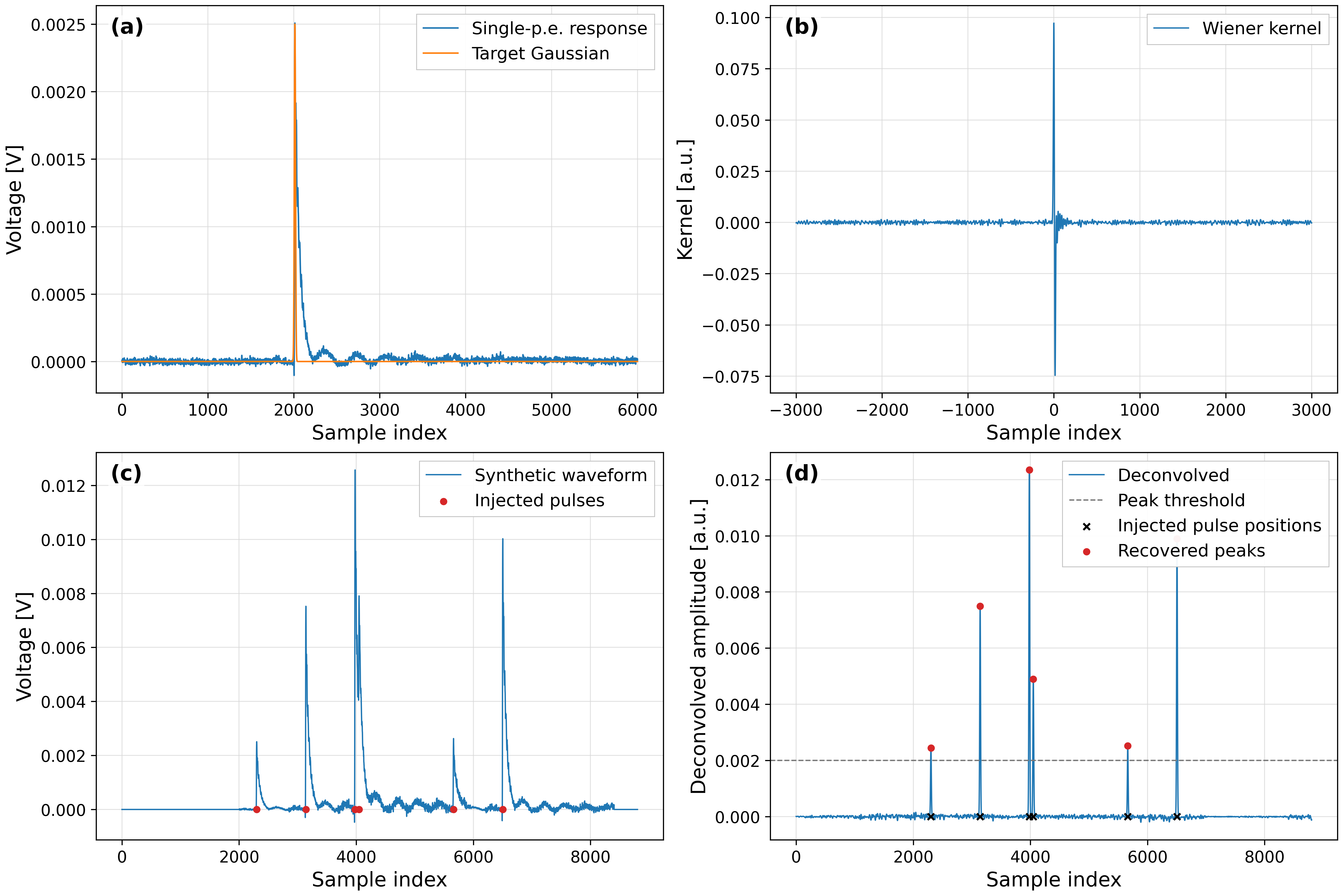}
    \caption{Validation of the Wiener deconvolution using synthetic pulses. (a) Measured single-p.e.\ response and target Gaussian. (b) Corresponding Wiener filter kernel. (c) Synthetic waveform constructed from pulses with known positions and amplitudes. (d) Deconvolved waveform showing the recovered peaks and peak-counting threshold.}
    \label{fig:deconv_validation}
\end{figure}

Here, $\delta[n]$ is used only to define the ideal inverse response. Direct application of $1/R[k]$, however, is unstable because frequency components where $R[k]$ is small can strongly amplify noise. Instead of reconstructing an ideally narrow delta function, the target response was chosen to be a narrow Gaussian d[n] defined as:

\begin{equation}
d[n] =
A_d
\exp\left[
-\frac{(n-n_0)^2}{2\sigma_d^2}
\right],
\end{equation}

where $n_0$ defines the pulse position, $A_d$ sets the output-amplitude scale, and $\sigma_d$ determines the pulse width. For this analysis, $\sigma_d=7$ samples and $A_d=0.0032$~a.u. were used. The value of $A_d$ defines the amplitude scale of the deconvolved waveform and does not represent the amplitude of the original SiPM flex pulse. The inverse-filter condition is then changed from a delta function to the target Gaussian:

\begin{equation}
(r*h_0)[n]=d[n],
\end{equation}

which in the frequency domain can be represented as: 

\begin{equation}
H_0[k]=\frac{D[k]}{R[k]},
\end{equation}

where $D[k]$ is the Fourier transform of the target Gaussian. The filter therefore removes the broad single-p.e.\ response and replaces it with the narrower Gaussian while preserving the pulse position. Figure~\ref{fig:deconv_validation}(a) shows the measured single-p.e.\ response and the target Gaussian used for this transformation. To control the noise amplification associated with the inverse response, the deconvolution was regularized using the measured noise power spectrum $P_N[k]$, defined in Eq.~\ref{eq:noise}. Following the Wiener-filter construction described in Ref.~\cite{HamamatsuMPPCGuide,Zhu2022Wiener}, the regularized filter was:

\begin{equation}
H_{\mathrm W}[k]
=
\underbrace{\frac{D[k]}{R[k]}}_{\text{target deconvolution}}
\underbrace{
\frac{|R[k]|^2}
{|R[k]|^2+\alpha P_N[k]+\epsilon}
}_{\text{noise suppression}},
\end{equation}

where $\alpha=10^{-4}$ controls the contribution of the measured noise spectrum and $\epsilon=10^{-12}$ prevents numerical instabilities when the denominator becomes very small. When the noise and regularization terms are negligible, $H_{\mathrm W}[k]$ reduces to the ideal Gaussian-target filter $H_0[k]=D[k]/R[k]$. The corresponding time-domain Wiener kernel, including the regularization term, is shown in Figure~\ref{fig:deconv_validation}(b). The time-domain kernel was obtained from:

\begin{equation}
h_{\mathrm W}[n]
=
\mathcal{F}^{-1}\!\left\{H_{\mathrm W}[k]\right\},
\end{equation}

and applied to each baseline-subtracted waveform as:

\begin{equation}
y_i[n]
=
\left(x_i^{\mathrm{corr}}*h_{\mathrm W}\right)[n],
\end{equation}

where $y_i[n]$ is the deconvolved waveform. The procedure was validated using synthetic waveforms constructed from measured single-p.e.\ responses with known pulse positions and amplitudes. Figure~\ref{fig:deconv_validation}(c) shows an example input waveform, while Figure~\ref{fig:deconv_validation}(d) shows the corresponding deconvolved waveform. The SiPM flex responses are reconstructed as narrow pulses at the injected positions, improving the separation of closely spaced signals and allowing their amplitudes to be compared with a common threshold for peak counting.

\subsubsection{Peak identification}

Peak finding was performed once on each deconvolved waveform using the minimum threshold of the scan. A sample $n$ was considered a peak candidate when the following conditions are met: 

\begin{equation}
y_i[n]\geq A_{\mathrm{min}},
\qquad
y_i[n]>y_i[n-1],
\qquad
y_i[n]\geq y_i[n+1].
\end{equation}

\begin{figure}[h!]
    \centering
    \includegraphics[width=0.96\linewidth]{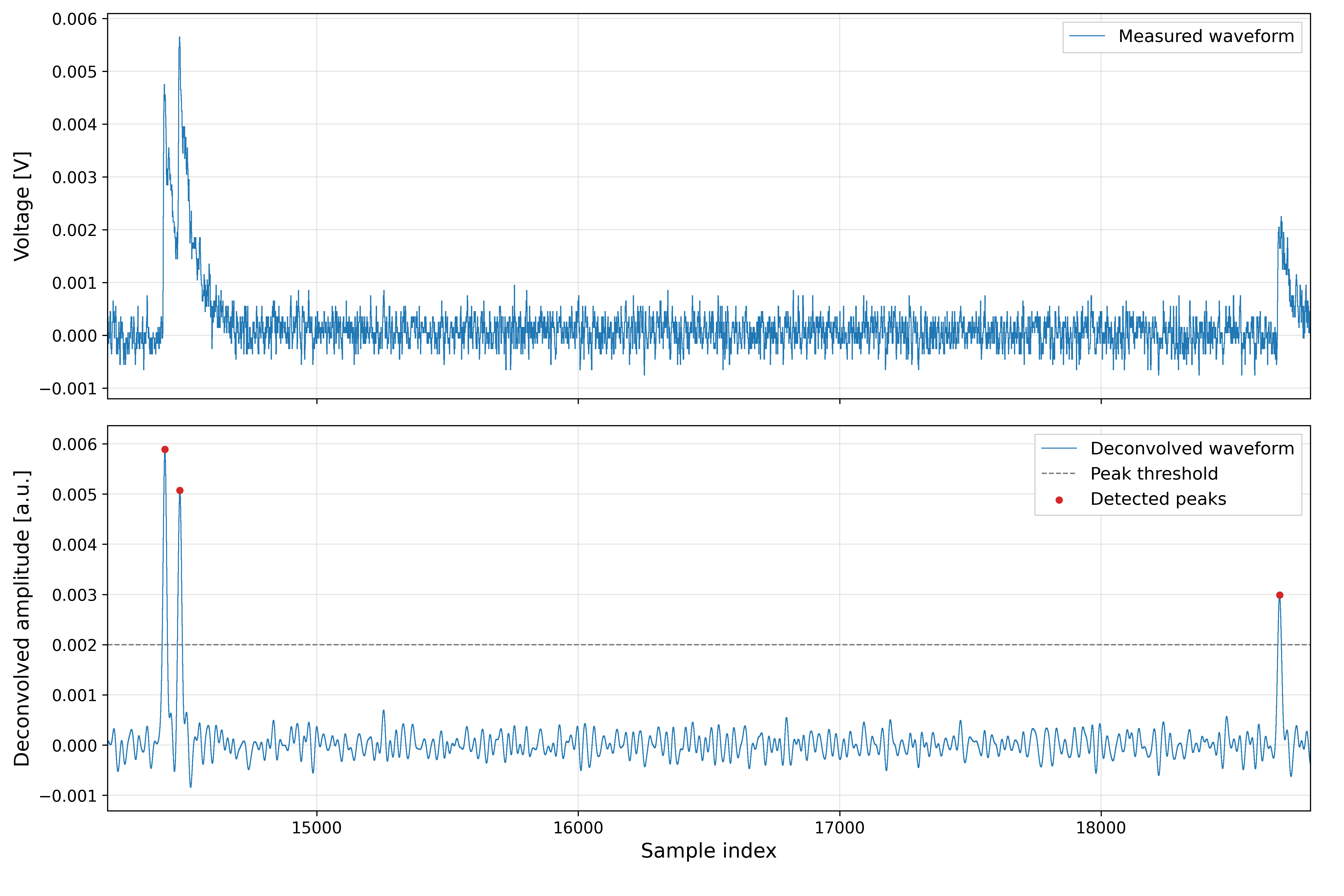}
    \caption{Example of peak identification in a measured SiPM flex waveform. The upper panel shows the recorded waveform, where nearby signals partially overlap. The lower panel shows the corresponding deconvolved waveform, where the signals are resolved into narrower peaks. The dashed line indicates the applied amplitude threshold, and the red markers indicate the peaks accepted by the identification procedure.}
    \label{fig:peak_resolution}
\end{figure}

Candidate peaks were ordered from largest to smallest amplitude. Beginning with the largest amplitude candidate, a peak was accepted only when it was separated from every previously accepted peak by at least the specified holdoff interval. This procedure retains the largest peak within each holdoff region and reduces double counting. For the data, $A_{\mathrm{min}}=0.0005$~a.u. and the holdoff interval was 20 samples, equivalent to 200~ns. After identifying the accepted peaks and their amplitudes, the same peak list was used for the complete threshold scan. At a deconvolved amplitude threshold $A_{\mathrm{thr}}$, the number of counted peaks was calculated as:

\begin{equation}
N_{\mathrm{peak}}(A_{\mathrm{thr}})
=
\sum_{p\in\mathcal{P}}
\Theta\!\left(A_p-A_{\mathrm{thr}}\right),
\end{equation}

where $\mathcal{P}$ is the set of accepted peaks, $A_p$ is the deconvolved amplitude of peak $p$, and $\Theta$ is the step function, such that $\Theta(A_p-A_{\mathrm{thr}})=1$ when $A_p\geq A_{\mathrm{thr}}$, and $\Theta(A_p-A_{\mathrm{thr}})=0$ when $A_p<A_{\mathrm{thr}}$. The deconvolved amplitude threshold $A_{\mathrm{thr}}$ was scanned from 0.0005 to 0.030~a.u. in steps of 0.0001~a.u. Figure~\ref{fig:peak_resolution} shows an example from the measured data in which nearby signals are more clearly resolved after deconvolution. The amplitude threshold and the peaks accepted by the identification procedure are also indicated.

\subsection{Results}

A separate dataset was acquired for each experimental configuration summarized in Tables \ref{tab:fiber_light_leakage} and \ref{tab:opc_light_leakage}. The fiber light leakage measurements were performed using the light-tight and non-light-tight fibers at wavelengths of 808, 1470, and 1550 nm. Both fibers were tested with one layer of PET mesh, while additional measurements were performed with the non-light-tight fiber without mesh at 1470 and 1550 nm. The OPC light leakage measurements were performed at 1470 and 1550 nm using both fiber types connected to their corresponding OPCs. The recorded SiPM flex waveforms were analyzed following the procedure described in Section \ref{sed:LKAnalysis}. For each dataset, the total live time was calculated as:

\begin{equation}
    T_{\mathrm{live}} =
    \sum_{i=1}^{N_{\mathrm{wf}}} \Delta t_i,
    \label{eq:live_time}
\end{equation}

where $N_{\mathrm{wf}}$ is the number of processed waveforms and $\Delta t_i$ is the duration of waveform $i$. Each dataset consisted of 3000 waveforms with a duration of 1~ms each, corresponding to a total live time of 3~s. The SiPM flex peak count rate was calculated as a function of the deconvolved amplitude threshold:

\begin{equation}
    R(A_{\mathrm{thr}}) =
    \frac{N_{\mathrm{peak}}(A_{\mathrm{thr}})}
    {T_{\mathrm{live}}}.
    \label{eq:count_rate}
\end{equation}

where $N_{\mathrm{peak}}(A_{\mathrm{thr}})$ is the total number of accepted peaks above the threshold and $T_{\mathrm{live}}$ is the total live time. The peak count rate $R(A_{\mathrm{thr}})$ was evaluated over the deconvolved amplitude threshold range rather than at a single threshold value. For each experimental configuration, an independent DCR measurement was acquired and analyzed using the same procedure. The resulting DCR peak count rates were then averaged to obtain a common DCR reference ($\overline{R}_{\mathrm{DCR}}$) for all datasets. This average DCR reference was subtracted from the peak count rate measured during PoF operation to determine the excess peak count rate associated with light leakage. \\

\begin{equation}
    R_{\mathrm{excess}}(A_{\mathrm{thr}}) =
    R_{\mathrm{PoF}}(A_{\mathrm{thr}})
    -
    \overline{R}_{\mathrm{DCR}}(A_{\mathrm{thr}}).
    \label{eq:excess_rate}
\end{equation}

\begin{figure}[h!]
    \centering
    \includegraphics[width=\linewidth]{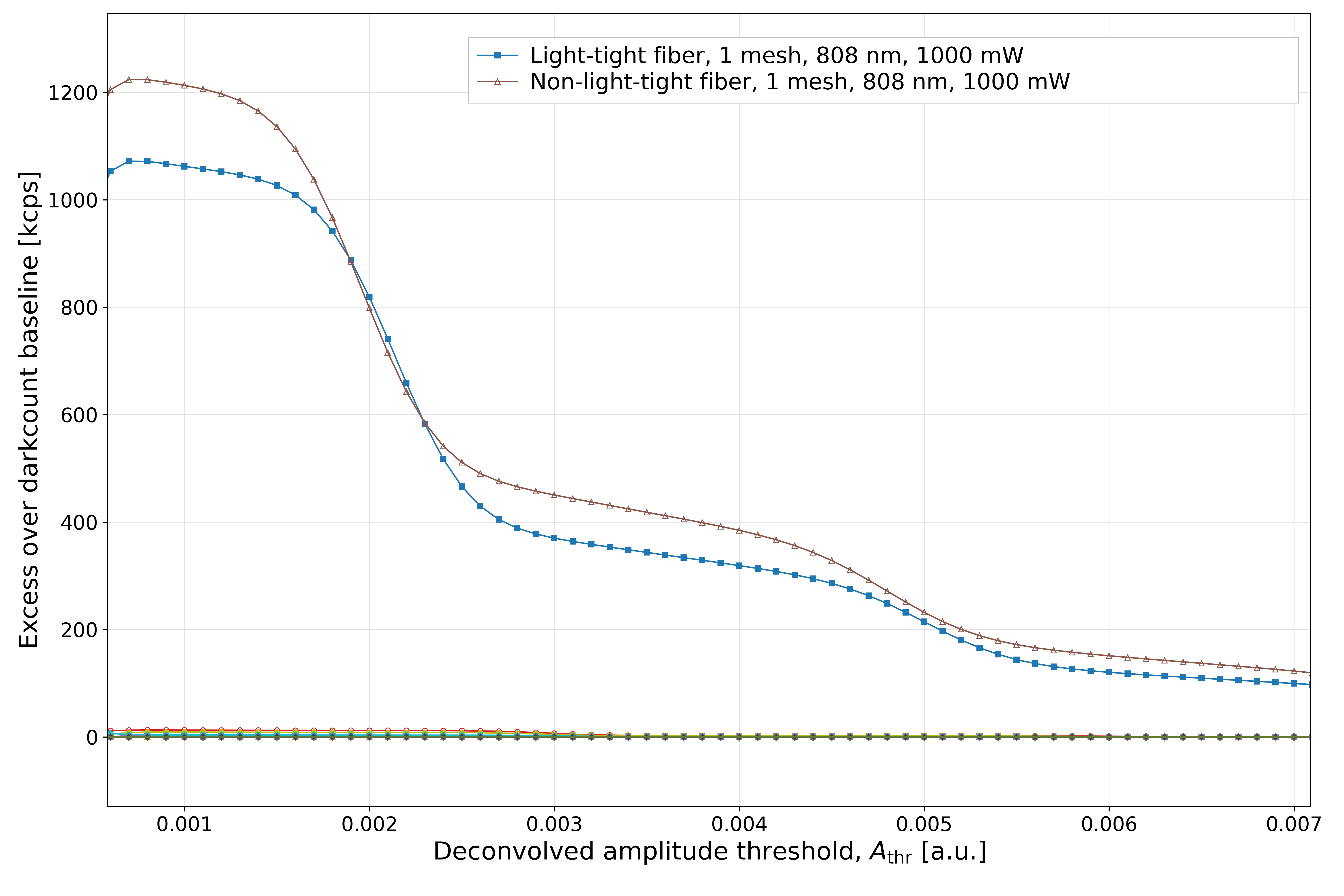}
    \caption{Excess SiPM flex count rate above the DCR baseline for the tested PoF configurations. The rate is shown as a function of the amplitude threshold applied to the deconvolved waveforms. The full-scale comparison includes the 808~nm reference measurement, while the curves near the baseline correspond to the 1470 and 1550~nm configurations.}
    \label{fig:exc1}
\end{figure}

Figure \ref{fig:exc1} shows the resulting excess SiPM flex count rates for the different experimental configurations. The measurements at 808 nm were used as a reference, since the SiPMs are sensitive to incident photons at this wavelength. For both the light-tight and non-light-tight fibers, a clear increase in the SiPM flex count rate above the DCR baseline was observed during 808 nm PoF operation. In comparison, the measurements at 1470 and 1550 nm show a reduction in the excess rate over the measured threshold range relative to the 808 nm measurements.

\begin{figure}[h!]
    \centering
    \includegraphics[width=\linewidth]{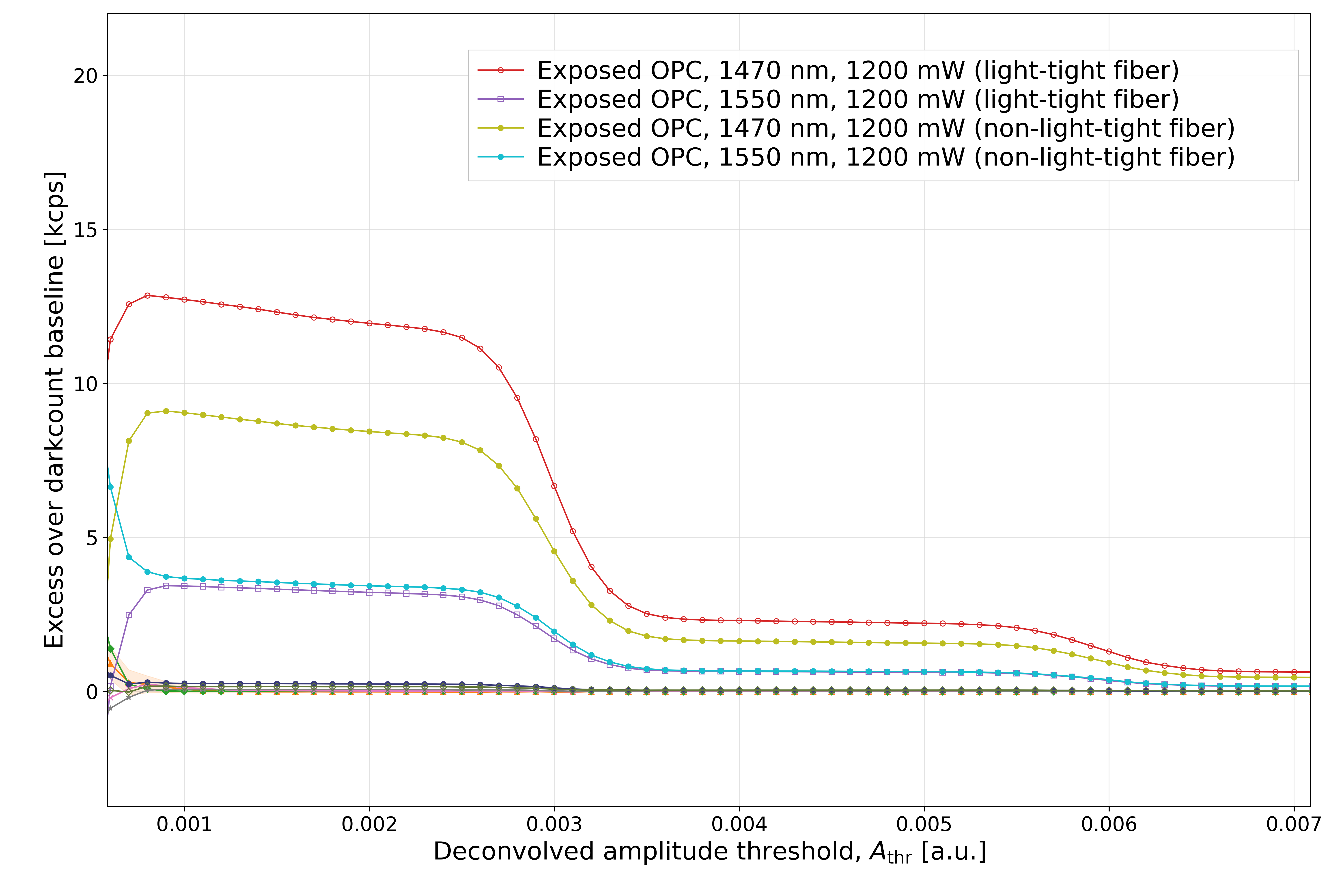}
    \caption{Expanded view of the excess SiPM flex peak count rate measured with exposed OPCs at 1470 and 1550~nm.}
    \label{fig:exc23_OPC}
\end{figure}

\begin{figure}[h!]
    \centering
    \includegraphics[width=\linewidth]{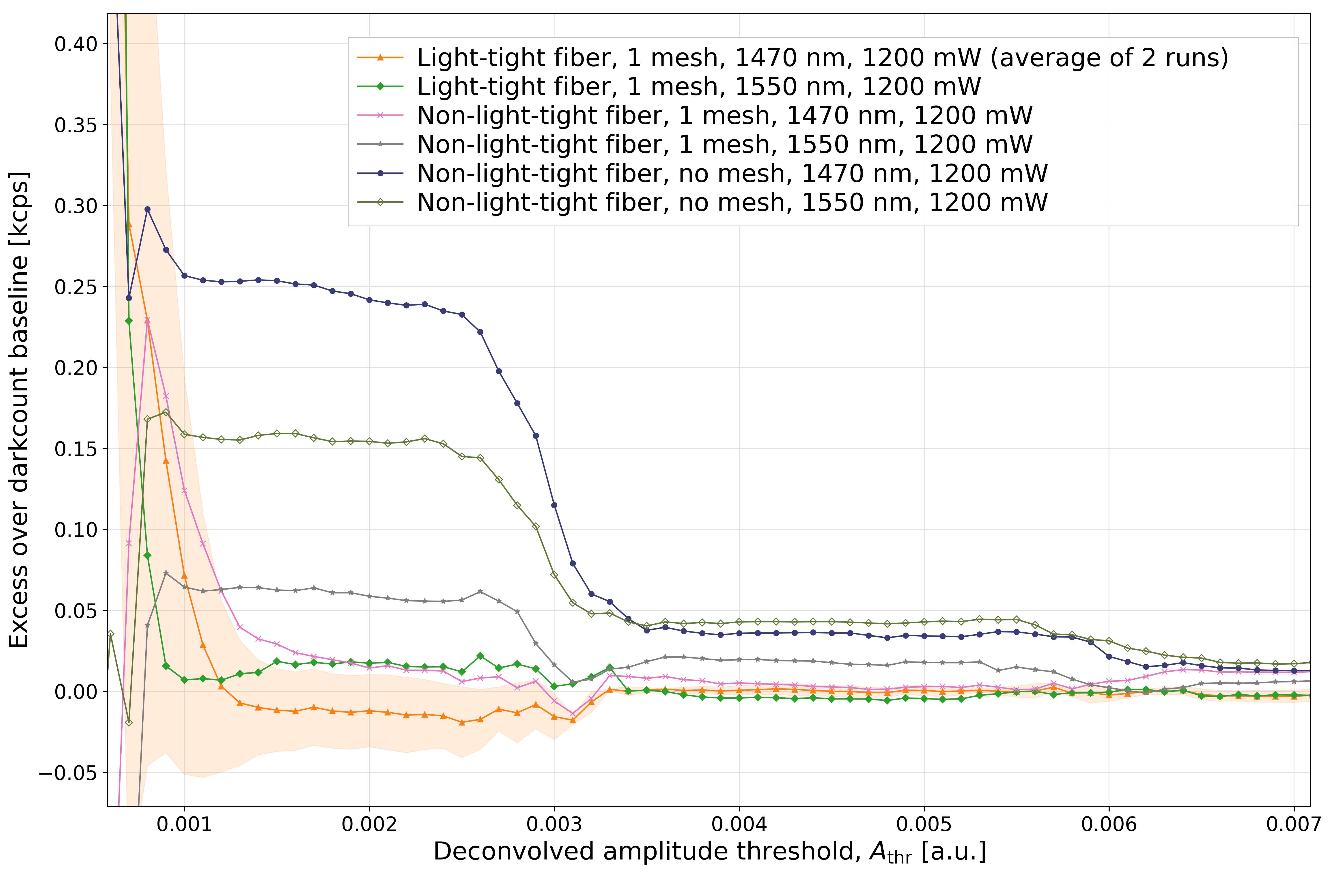}
    \caption{Expanded view of the excess SiPM flex count rate measured at 1470 and 1550~nm for the fiber-only configurations, with and without protective PET mesh.}
    \label{fig:exc23_fiberonly}
\end{figure}

Due to the large difference between the 808 nm reference and the 1470 and 1550~nm wavelength measurements, the latter appear close to the baseline in Figure \ref{fig:exc1}. To visualize a more detailed comparison, Figures \ref{fig:exc23_OPC} and \ref{fig:exc23_fiberonly} show the results at 1470 and 1550 nm using reduced vertical-axis ranges for the OPC and fiber-only configurations, respectively.

Figure \ref{fig:exc23_OPC} shows the excess count rates measured with the OPC connected to the optical fiber. In this configuration, the OPC and $\sim$10 cm of the connected fiber were placed inside the Dewar, while the remaining fiber was kept outside. The measurements with the exposed OPC showed higher excess rates compared with the fiber-only configurations, reaching up to approximately 13 kcps. These rates correspond to roughly 15 to 50 counts in a 4 ms window. The selection of this time window was motivated by the DUNE TDR PDS simulations \cite{DUNEFDVD2024}.

Figure \ref{fig:exc23_fiberonly} shows the excess count rates measured for the optical fibers with and without PET mesh. At both 1470 and 1550 nm, the excess rates remained close to the DCR baseline, generally below approximately 0.3 kcps over most of the threshold range. This corresponds to approximately one count or less within a 4 ms acquisition window. 

These results highlight the importance of developing next-generation PoF systems operating at longer wavelengths to power the cold readout electronics of future large-area photon detectors operating in harsh environments, such as the proposed photon detection system for DUNE Phase II \cite{Marinho2025APEX,DUNE2024PhaseII}.
\section{Future applications of next generation PoF system}
\label{sec:applications}
\subsection{Applications in large area photon detectors}
The PoF system being developed for the DUNE Vertical Drift (VD) photon detection system can serve as a reference for future HEP detectors where electrical isolation, low-noise, and a larger number of photon-detector units are required. One example is the Aluminum Profiles with Embedded X-ARAPUCA (APEX) concept proposed for the DUNE Phase II third far-detector module~\cite{Marinho2025APEX,DUNE2024PhaseII}. In APEX, photon-detector units would be integrated into the four vertical field-cage walls, providing up to approximately 60\% optical coverage with about 7000 units distributed over these walls~\cite{Marinho2025APEX}. Because the readout electronics need to operate at the field-cage potential, power and signal connections must remain electrically isolated. PoF for power delivery and Signal-over-Fiber (SoF) for signal transmission are being considered for this detector concept~\cite{DUNE2024PhaseII,DUNE2025PhaseIIDetectors,Marinho2025APEX}, building on the experience gained for the preparation of the DUNE VD PDS ~\cite{DUNEFDVD2024}.

Scaling PoF to detectors with larger optical coverage increases the relevance of a next generation PoF system, particularly in terms of light leakage reduction, power-conversion efficiency, and fiber integration. The measurements presented in Section \ref{sec:light_leak} at 1470 and 1550~nm show a reduction of approximately two orders of magnitude in the excess SiPM count rates compared with the 808~nm PoF system under the tested conditions. Beyond light leakage, the electrical power available after optical transmission is an important consideration for PoF systems operating in cryogenic environments. The fiber attenuation coefficients and OPC optical-to-electrical power conversion efficiencies presented in Sections \ref{sec:opticalfibers} and \ref{sec:OPC} can be combined to estimate the available electrical power for different fiber lengths and laser transmitter optical-power settings. For the characterized fiber and OPCs in this paper, the projected electrical OPC output power is estimated as:

\begin{equation}
    P_{\mathrm{out}}(L)
    =
    \eta_{\mathrm{OPC}}P_{\mathrm{laser}}
    10^{-[A_0+(\alpha_{\mathrm{fiber,dB}}+\alpha_{\mathrm{cryo,dB}})L]/10},
    \label{eq:opc_power_projection}
\end{equation}

where $\eta_{\mathrm{OPC}}$ is the measured optical-to-electrical conversion efficiency, $P_{\mathrm{laser}}$ is the laser transmitter optical output power, $A_0$ is the fixed insertion loss, and $\alpha_{\mathrm{fiber,dB}}$ and $\alpha_{\mathrm{cryo,dB}}$ are the room-temperature and induced cryogenic attenuation coefficients, respectively, expressed in dB per unit length. These projections assume that the measured attenuation coefficients and OPC conversion efficiencies remain applicable over the considered fiber lengths and laser transmitter optical-power range.

\begin{figure}[!h]
    \centering
    \includegraphics[width=0.95\linewidth]{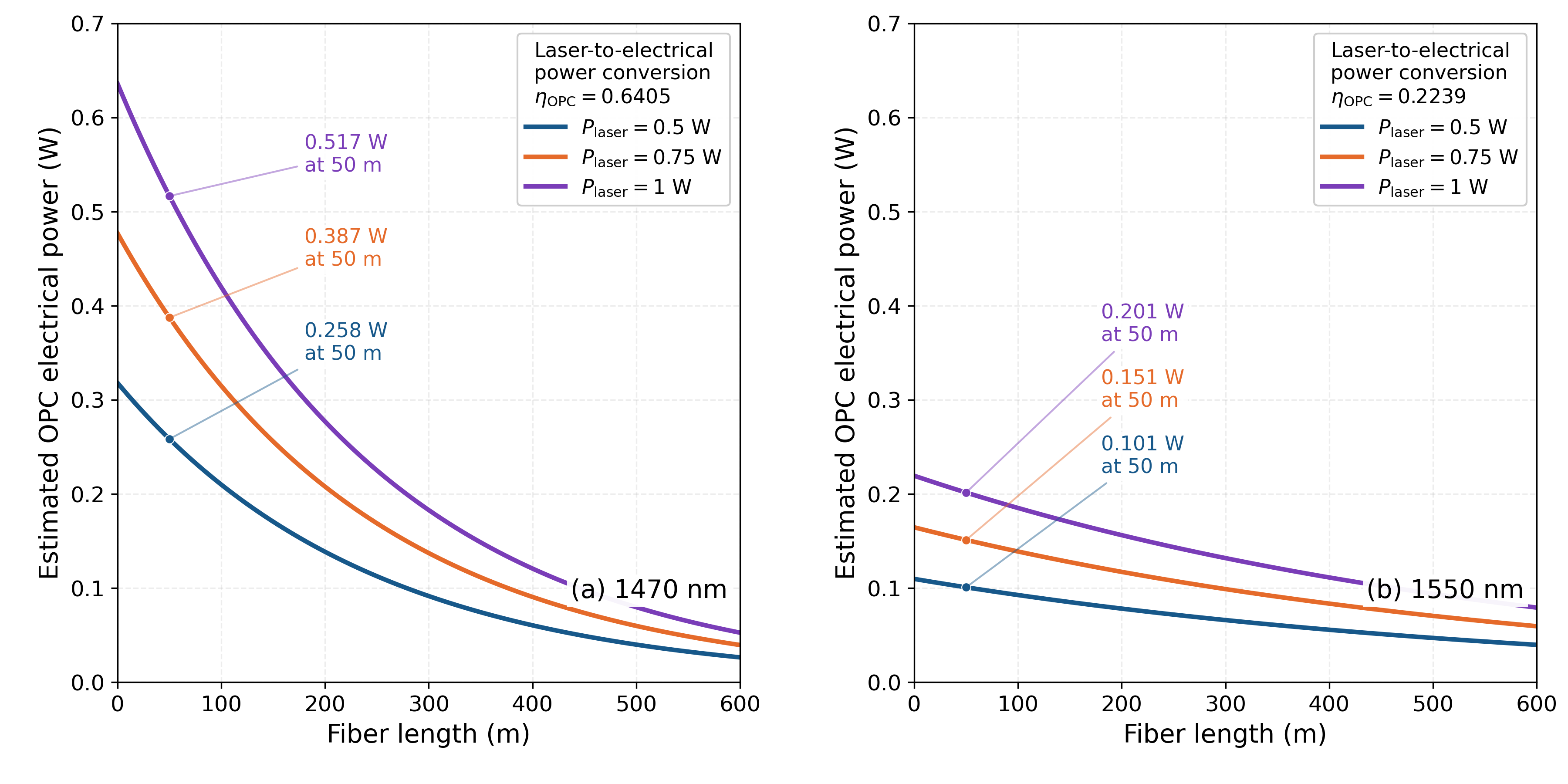}
    \caption{Projected electrical output power from the OPC as a function of fiber length at (a) 1470~nm and (b) 1550~nm for three laser optical-power settings. The curves combine the estimated fiber transmission losses with the measured OPC conversion efficiencies at cryogenic temperatures. The values indicated at 50~m represent the projected electrical power available at the OPC output for each setting.}
    \label{fig:opc_power_projection}
\end{figure}

Figure~\ref{fig:opc_power_projection} shows the projected OPC electrical output power for three laser optical-power settings at 1470 and 1550~nm. A fiber length of approximately 50~m was considered as a conservative reference for a PoF system with future detector applications, including the proposed APEX system. At this length, for example, a laser optical output power of 1~W is projected to deliver approximately 0.517~W of electrical power at 1470~nm and 0.201~W at 1550~nm. These results illustrate how the measured parameters can be used to design a PoF system that provide an electrical power for a selected fiber length, OPC, and laser transmitter operating condition. 

Further tests are required before extending the new PoF system to larger photon detectors such as the proposed APEX PDS for DUNE Phase II. These could include determining the power requirements of the complete photon-detector cold readout electronics, defining the number and routing of optical power fibers, and studying long-term operation of the full PoF-PDS chain in cryogenic conditions.  

\subsection{Applications in other research fields}

Optical fibers have been used in industrial systems where their electrical isolation and ability to cover long distances provide advantages over conventional conductive connections. A common example of this are electrical substations, where fiber-optic systems have been implemented for monitoring equipment \cite{Rosolem17}. Remote telecommunications equipment is also an application of PoF systems, where optical power can be delivered through the same fiber infrastructure used for data transmission \cite{perez2026power}. Similarly, PoF systems has been used to operate remote gas sensors in mines, where the power and sensor data are sent simultaneously \cite{chilton1995fiber}, and also in applications for environmental and geotechnical monitoring \cite{rodhiah2025development}. 
Properties of optical fibers make them a useful choice for systems operating at cryogenic temperatures, where the thermal and electrical conductivity of conventional metallic cables can impose significant constraints. This could be relevant for superconducting quantum computing, in which large numbers of electrical connections between room-temperature electronics and cryogenic circuits could produce unwanted heat loads. Tests of optical fibers for data transmission in these systems are promising, with very low heat transfers while maintaining a long link distance. The characterization of the PoF system components described in this paper highlight potential use cases in different research fields.
As shown in Section \ref{sec:OPC}, high efficiency OPCs operating at 1470 nm in cryogenic environments present an optical-to-electrical power conversion efficiency of ~64$\%$. Combining this efficiency with the measured optical fiber attenuation at room and cryogenic conditions for 1470 nm (see Section \ref{sec:opticalfibers}) these results could motivate the design and operation of remote systems in harsh environments using PoF. Additionally, Section \ref{sec:light_leak} presents the light leakage results and comparisons between PoF systems operating at different wavelengths. This could also present further opportunities for work in the quantum computing and sensing sectors, where unwanted light may cause quantum decoherence. Furthermore for applications in environments exposed to extreme radiation, such as instrumentation in space, fission reactors, or in radiative material processing where the fiber is exposed to heavy ionizing radiation could cause radiation-induced attenuation in the fiber. Depending on the optical fiber composition and irradiation conditions, radiation-induced attenuation can be higher at shorter wavelengths, such as 808~nm, than at wavelengths near 1470 and 1550~nm~\cite{girard2013radiation}.
This, along with the lower weight and volume of optical fibers compared to the equivalent electrical connections, makes optical fiber technologies operating at these wavelengths an attractive option for operation in this type of harsh environment.
\section{Conclusions}
\label{sec:summary}
This paper presents a characterization of each of the components of the next generation PoF system. Off-the-shelf lasers of operating wavelengths 1470 and 1550~nm  were characterized and we measured a maximum electrical-to-optical conversion efficiencies of $\sim$24\% and $\sim$28\% respectively. Additionally, preliminary results of a customized laser driver prototype capable of operating up to ~7A while delivering a maximum optical power of ~2.6 W were presented. The laser driver prototype was also operated during one hour with a current of 3 A and was able to deliver up to 1 W of optical power with optical power fluctuations below 1\%. A quantification of the attenuation in the optical fibers studied in this paper in both room and cryogenic temperatures resulted in the obtaining the following attenuation coefficients: 
\begin{table}[h!]
    \centering
    \begin{tabular}{lcc}
        Parameter & 1470 nm & 1550 nm \\
        \hline
        $A_0$ [dB]
        & $0.03 \pm 0.01$
        & $0.09 \pm 0.02$ \\
        $\alpha_{\mathrm{fiber,dB}}$ [dB/km]
        & $13.28 \pm 0.96$
        & $1.38 \pm 1.03$ \\
        $\alpha_{\mathrm{cryo,dB}}$ [dB/km]
        & $4.77 \pm 0.31$
        & $5.99 \pm 0.30$ \\
        $\alpha_{\mathrm{combined,dB}}$ [dB/km]
        & $18.05 \pm 1.01$
        & $7.37 \pm 1.07$ \\
    \end{tabular}
\end{table}

Measurements of the optical-to-electrical conversion efficiencies resulted in a $\eta = 0.6405 \pm 0.0324$ for the 1470~nm OPC and $\eta = 0.2239 \pm 0.0118$ for the 1550~nm OPC. The use of the 1470~nm OPC in PoF systems operating under cryogenic conditions represents an increase in optical-to-electrical conversion efficiency of approximately 10 percentage points compared with the OPCs used in 808~nm PoF systems. The measurement of the optimal load resistance for both 1470 nm and 1550 nm OPCs to reach the maximum power followed a model with inverse dependence of the OPC input power. 

Finally, the response of Broadcom SiPMs to IR photons from optical fiber leakage was characterized. The measurements included multiple configurations of optical fibers, protective meshes, and OPCs. The measurements show that the excess SiPM count rates associated with PoF operation at 1470 and 1550 nm were reduced by approximately two orders of magnitude or more compared with the 808 nm reference under the tested conditions. The exposed OPC configurations produced excess count rates reaching approximately 13 kcps, while the fiber-only configurations showed excess rates below 0.3 kcps. 
Altogether, these results serve as the foundation to the design of a PoF system for powering proposed large area photon detectors, such as APEX in DUNE Phase II, and other applications where remote devices are needed for operation in harsh environments. Future deployment of the next generation PoF system presented in this paper could take place at CERN in DUNE Phase II prototypes before the end of this decade. The procedures developed for characterizing the PoF components presented in this paper, particularly the estimation of fiber attenuation at cryogenic temperatures, could be applied in other research fields, such as space exploration and remote system monitoring, where power must be transmitted over long distances. These methods could help in selecting suitable optical fibers for cryogenic operation and, together with the light-leakage measurements, motivate further studies of PoF applications for powering electronics in cryogenic environments, including quantum devices that may be sensitive to infrared photons.

\acknowledgments
The authors would like to thank Dante Totani for his valuable discussions and technical feedback regarding SiPM signal readout and light-leakage. The authors also gratefully acknowledge to Simon Fafard, Jan Werthen, Denis Masson, and Terry Schaul from Broadcom for their technical discussions including feedback on the OPCs studied in this paper. Work at South Dakota School of Mines and Technology has been supported by funding under the U.S. Department of Energy, Office of Science, Office of High Energy Physics, Early Career Research Program Award Number DE-SC0025587. The manuscript has been authored by Fermi Forward Discovery Group, LLC under Contract No. 892430-\hspace{0pt}24CSC00\hspace{0pt}0002 with the U.S. Department of Energy, Office of Science, Office of High Energy Physics. Publisher acknowledges the U.S. Government license to provide public access under the DOE Public Access Plan. 

\bibliographystyle{JHEP}
\bibliography{biblio}

\end{document}